\documentclass[journal]{IEEEtran}

\IEEEoverridecommandlockouts

\usepackage[utf8]{inputenc} 

\usepackage[T1]{fontenc}
\usepackage{url}
\usepackage{ifthen}
\usepackage{cite}
\usepackage{overpic}
\usepackage[cmex10]{amsmath} 

\usepackage{textcase}
\usepackage[tablename=Table]{caption}
\usepackage{blindtext}
\usepackage{floatrow}
\usepackage{soul}
\usepackage{changes}
\usepackage{gensymb}
\usepackage{cite}
\usepackage{amsmath,amssymb,amsfonts}

\usepackage{textcomp}
\usepackage{graphicx}

\usepackage{amssymb}
\usepackage{amsfonts}
\usepackage{amsmath}
\usepackage{epsfig}
\usepackage{epigraph}
\usepackage{color}
\usepackage{fancybox}
\usepackage{textcomp}
\usepackage{multirow}
\usepackage{setspace}
\usepackage{psfrag}
\usepackage{booktabs}
\usepackage{float}
\usepackage[caption = false]{subfig}
\usepackage{algorithm}
\usepackage{algpseudocode}
\usepackage{mathtools, nccmath, bigints, amsfonts}
\usepackage{array}
\usepackage{stfloats}
\newcolumntype{L}{>{\centering\arraybackslash}m{3cm}}

\usepackage{amsthm}
\theoremstyle{definition}

\usepackage{mathrsfs}

\newfloatcommand{capbtabbox}{table}[][0.4\textwidth]
 \usepackage[font={small},figurename=Fig.]{caption}
\newtheorem{Theorem}{Theorem}
\newtheorem{corollary}{Corollary}
\newtheorem{Lemma}{Lemma}

\makeatletter 
\pretocmd\@bibitem{\color{black}\csname keycolor#1\endcsname}{}{\fail}
\newcommand\citecolor[1]{\@namedef{keycolor#1}{\color{black}}}
\makeatother
\citecolor{irregular2017}
\citecolor{bookEmil}
\citecolor{degreesoffreedom}
\citecolor{Yuan_2024}
\citecolor{irregularuca24}
\citecolor{multiusertri}
\citecolor{zhu2024}
\citecolor{moveable23}
\citecolor{CAVALIERE201756}

    \def\Complex{{\rm\rule[.23ex]{.03em}{1.1ex}\kern-.3em{C}}}

    \newcommand{\be}{\begin{equation}} \newcommand{\ee}{\end{equation}}
    \newcommand{\bea}{\begin{eqnarray}} \newcommand{\eea}{\end{eqnarray}}
    \newcommand{\benum}{\begin{enumerate}} \newcommand{\eenum}{\end{enumerate}}

\def\imaginary{\mathsf{j}} 
\def\Htran{\mbox{\tiny $\mathrm{H}$}}
\def\Ttran{\mbox{\tiny $\mathrm{T}$}}
\begin{document}

\title{Optimal Dual-Polarized Planar Arrays for Massive Capacity Over Point-to-Point MIMO Channels}

\author{Amna~Irshad,~Alva~Kosasih,~Emil~Bj{\"o}rnson,~\IEEEmembership{Fellow,~IEEE},~Luca~Sanguinetti,~\IEEEmembership{Senior Member,~IEEE}  \\
\thanks{A preliminary version of this article was presented at the European Signal Processing Conference (EUSIPCO) 2023 \cite{irshad2023optimal}. A. Irshad, A. Kosasih, and E. Bj{\"o}rnson are with the Division of Communication Systems,  KTH Royal Institute of Technology, SE-100 44 Stockholm, Sweden. Email: \{amnai,kosasih,emilbjo\}@kth.se.}%
\thanks{L. Sanguinetti is with the Dipartimento di Ingegneria dell'Informazione, University of Pisa, 56122 Pisa, Italy. Email: luca.sanguinetti@unipi.it}%
\thanks{This paper was supported by the Grants 2019-05068 and 2022-04222 from the Swedish Research Council. L. Sanguinetti was partially supported by the Italian Ministry of Education and Research (MUR) in the framework of the FoReLab project (Departments of Excellence).
}}

\maketitle

\begin{abstract}
Future wireless networks must provide ever higher data rates. The available bandwidth increases roughly linearly as we increase the carrier frequency, but the range shrinks drastically.
This paper explores if we can instead reach massive capacities using spatial multiplexing over multiple-input multiple-output (MIMO) channels. In line-of-sight (LOS) scenarios, the rank of the MIMO channel matrix depends on the polarization and antenna arrangement.
We optimize the rank and condition number by identifying the optimal antenna spacing in dual-polarized planar antenna arrays with imperfect isolation. The result is sparsely spaced antenna arrays that exploit radiative near-field properties.
We further optimize the array geometry for minimum aperture length and aperture area, which leads to different configurations.
Moreover, we prove analytically that for fixed-sized arrays, the MIMO rank grows quadratically with the carrier frequency in LOS scenarios, if the antennas are appropriately designed. Hence, MIMO technology contributes more to the capacity growth than the bandwidth.
The numerical results show that massive data rates, far beyond 1 Tbps, can be reached both over fixed and mobile point-to-point links. It is also possible for a large base station to serve a practically-sized mobile device.

\end{abstract}

\begin{IEEEkeywords}
Line-of-Sight MIMO, Dual-Polarized, Optimal Capacity, Near-Field, Terahertz frequencies.
\end{IEEEkeywords}

\section{Introduction}

The capacity requirements on wireless communication links continue to grow, and we cannot cater to them by indefinitely increasing the spectral bandwidth, which is a limited resource. Alternatively, multiple-input multiple-output (MIMO) technology can be used to improve capacity. MIMO technology enables spatial multiplexing in rich multi-path propagation scenarios \cite{Telatar1999a}, where 
the capacity and rank of the MIMO channel are proportional. 

The channel capacity can be expressed as 
${\rm rank(\mathbf{H})}\log_2(\rm SNR)$ at high signal-to-noise ratio (SNR) \cite{degreesoffreedom}, where $\mathbf{H}$ denotes the $M \times K$ channel matrix with $M$ outputs and $K$ inputs. This means that the capacity growth depends linearly on the channel rank,  referred to as the degrees of freedom (DOF)\cite{bookEmil}. The maximum DOF of $\min(M,K)$ is achieved when the channel matrix has full rank \cite{Yuan_2024}, demonstrating that full-rank channels are desirable in MIMO communications.  

However, sixth-generation (6G) mobile systems operating at mmWave and sub-terahertz frequencies will feature line-of-sight (LOS) dominant channel conditions \cite{Rappaport2019}. Free-space point-to-point LOS MIMO channels were traditionally viewed to have rank $1$ \cite[Sec.~7.2.3]{Tse2005a} since only a single planar wave can be transferred from the transmitter's antenna array to the receiver's array.
However, a higher rank can be achieved by capitalizing on the radiative near-field region of the electromagnetic field, as discussed in \cite{Lu2023}. This is attributed to the spherical wavefront within the radiative near-field region, which can be utilized to enhance the LOS MIMO rank, as discussed in \cite{Decarli2021a}.
The demarcation line between the near-field and far-field regions is based on the Fraunhofer distance,  given by $d_{\mathrm{FA}} = \frac{2D^2}{\lambda}$ (e.g., \cite{Ramezani2023b}) where $D$ is the aperture length and $\lambda$ the wavelength. This becomes particularly relevant with the utilization of the high frequencies envisioned in 6G, as it significantly extends the Fraunhofer distance, owing to its inverse proportionality to the wavelength.
For example, the Fraunhofer distance for a large array operating at $f=3\,$GHz with size $\sqrt 2  \times \sqrt 2$\,m and aperture length $D = 2\,$m is $d_{\mathrm{FA}} = 80\,$m. If the same array operates at $f = 30\,$GHz, the Fraunhofer array distance increases to $d_{\mathrm{FA}} = 800\,$m. If new deployment practices enable even larger arrays, then the Fraunhofer distance can extend to several kilometers. Consequently, a significant number of potential users served by the base station would fall within the near-field region, where the spherical wavefront is prominent.

This paper explores how to maximize the capacity of point-to-point LOS MIMO channels by optimizing the antenna array geometries, under size constraints and for varying carrier frequencies. Apart from establishing fundamental insights, the results are useful for designing high-capacity wireless backhaul/fronthaul links and fixed wireless access links, which are two classical use cases that will remain important in future network generations \cite{CAVALIERE201756}. The results also have implications for the array design in mobile user devices. Cellular networks must nowadays use multi-user MIMO protocols to achieve large DOF since the individual user channels have low rank. However, if we can optimize the arrays to achieve high-rank user channels, then point-to-point MIMO with orthogonal user scheduling becomes a viable alternative way to achieve high capacity in future networks. This is particularly important when operating systems at mmWave and sub-terahertz frequencies, where the range is short and the number of simultaneously active users can be small. Moreover, cellular networks can use point-to-point MIMO in low-traffic hours, while multi-user MIMO requires multiple active users to be effective.

\subsection{Related Works}

The maximum rank that a continuous \emph{holographic} array of a given size can achieve was characterized (among others) in \cite{Hu2018a} and \cite{Pizzo2020a}. While this characterization is not directly applicable to LOS MIMO communication links, it serves as a bound on the maximum rank in MIMO communications. It is also worth noting that not only the rank determines the channel capacity but also the condition number of the channel matrix, which should ideally be one.
Considering an ideal LOS MIMO scenario, it was shown in \cite{Bohagen2007a,multi2011,Do2021a} how the antenna spacing in two uniform linear arrays (ULAs) can be optimized to achieve both a rank equal to the number of antennas and a condition number of one for the channel matrix (which is the ideal case for spatial multiplexing). The aforementioned papers consider arrays with parallel broadside directions, while the rotation of antenna arrays was studied in \cite{Jeng2005,Garca2018LOSMD}. These papers also optimized the antenna spacing and showed that the spherical wave model is more accurate than the planar in MIMO system. 
In \cite{renzo2023}, the spherical wavefront was exploited to design the uniform linear arrays for efficient performance and is compared with conventional half-wavelength spaced arrays. The accuracy and relevance of the parabolic model in near-field MIMO is investigated in \cite{Heedong2023}.
The results were extended to the case of uniform rectangular arrays (URAs) also termed two-dimensional arrays in \cite{LarssonP.2005,Zhou2012AADf}. The optimization of the singular values in the channel matrix for uniform circular arrays (UCAs) is discussed in \cite{irregularuca24} and for URAs in \cite{Bohagen2007b}, and ULA was observed as a special case of URA. The work in \cite{irregular2017} discussed capacity for irregular arrays motivating the usage more inclined towards reconfigurable antenna designs. However, these prior works are limited to single-polarized arrays, although practical systems generally utilize dual polarization. The dual-polarized antenna allows us to send parallel streams in the same signal direction, doubling the capacity, when there is no signal leakage or radiations into the unintended polarization direction.
This is captured by the cross-polar discrimination (XPD). To the best of our knowledge, this is the first time that dual-polarized antennas have been considered when deriving the optimal spacing in point-to-point MIMO systems.

\subsection{Contributions}

In this paper, we consider a LOS MIMO channel between two dual-polarized URAs, with imperfect isolation. 
 
We first derive the antenna spacing for the URAs that optimizes the eigenvalues of the channel matrix, consequently maximizing the system's capacity. Then, we minimize the physical aperture length and area while retaining the maximum capacity. Finally, we investigate the behavior of system capacity in relation to the utilization of higher carrier frequency bands.  The main contributions are as follows.
\begin{itemize}
    \item We analytically derive the horizontal and vertical antenna spacing that maximizes the MIMO channel capacity in the spatial multiplexing regime where the signal-to-noise ratio (SNR) is large. The results are qualitatively and quantitatively different than in the prior work due to imperfect polarization isolation \cite{LarssonP.2005,Zhou2012AADf}. 
    \item We utilize the new analytic results to optimize the array geometries further to minimize either the area of the URA or the maximum aperture length. The analytical results are corroborated numerically.
    \item The antenna gain for both transmitter and receiver array is optimized in terms of directivity to attain the maximum capacity by moving towards higher frequencies.
\end{itemize}

\subsection{Paper outline and notation}
The paper is organized as follows.  Section~\ref{Sect_Syst_Mod} introduces the system model, i.e., the LOS channel between two dual-polarized arrays. Section~\ref{Sect_Capacity_max} derives the antenna spacing maximizing the system capacity with URAs. The aperture length and area minimization problem are also discussed.  Section~\ref{Sect_Capacity_Freq} analyzes the capacity as the carrier frequency and thus the bandwidth increase. Section~\ref{Sect_conclusion} concludes the paper.

Bold uppercase letters are used to identify matrices. For a matrix $\mathbf{A}$, $\mathbf{I}_{2}$ represents $2 \times 2$ identity matrix. The $\mathbf{A}$ $\otimes$ $\mathbf{B}$ represents the Kronecker product. For two variables: $a$ and $b$, a is defined to be equal to be and is represented as $a \triangleq b$.

\section{System Model}\label{Sect_Syst_Mod}

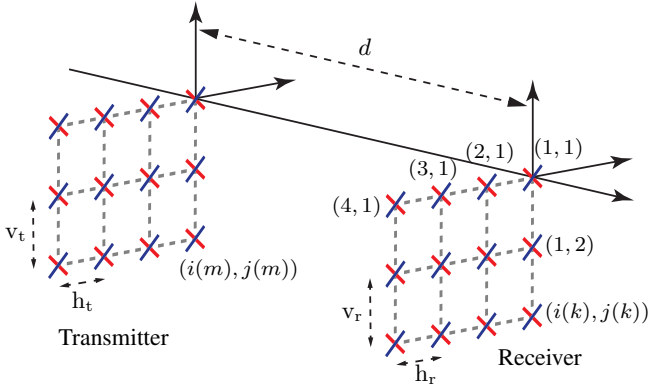
\begin{figure}[t!]
			\begin{overpic}[width=\columnwidth,tics=10]{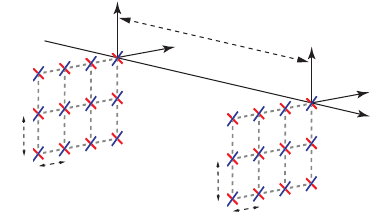}
                \put(12,10.5){\small ${\rm h}_{\rm t}$}
                \put(28,15.5){\footnotesize $(i(m),j(m))$}
				\put(2,21){\small ${\rm v}_{\rm t}$}
    		    \put(55,48.5){\small $d$}
    		    \put(10,5){\small Transmitter}
    		    \put(76,2){\small Receiver}
				\put(51,25){\footnotesize $(4,1)$}
				\put(83,19){\footnotesize $(1,2)$}
				\put(83,9){\footnotesize $(i(k),j(k))$}
                \put(63.5,-0.5){\small ${\rm h}_{\rm r}$}
                \put(52.5,9){\small ${\rm v}_{\rm r}$}
				\put(62.5,30.5){\footnotesize $(3,1)$}
				\put(71,33){\footnotesize $(2,1)$}
				\put(81.3,33.7){\footnotesize $(1,1)$}
			\end{overpic} 
\caption{ A LOS channel between two dual-polarized arrays separated by a distance $d$. In this example, $M_{\rm h}=4$ and $M_{\rm v}=3$, while the horizontal/vertical indices are shown on the receiver side. The antenna spacings ${\rm v}_{\rm t},{\rm h}_{\rm t},{\rm v}_{\rm r},{\rm h}_{\rm r}$ will be optimized to maximize capacity.}
\label{figsystemmodel}
\end{figure}

We consider a  point-to-point free-space LOS channel between two planar arrays, separated by a distance $d$. The arrays are aligned in broadside directions and arranged as URAs with $M_{\rm  v}$ vertically stacked rows  and $M_{\rm  h}$ antennas per horizontal row. This makes the total number of antenna locations $M=M_{\rm  v}M_{\rm  h}$. The vertical and horizontal spacings are respectively denoted as ${\rm v}_{\rm t}$ and ${\rm h}_{\rm t}$ at the transmitter, and ${\rm v}_{\rm r}$ and ${\rm h}_{\rm r}$ at the receiver. These spacings will later be optimized. Each antenna is dual-polarized, and consists of two co-located elements with orthogonal polarization dimensions (e.g., slanted $\pm 45^\circ$) \cite{Emil2017}. Hence, the total number of antenna elements per array is $2M$.
The antenna locations in each array are numbered row by row from $1$ to $M$, and for a given antenna index $m \in \{ 1,\ldots,M\}$, the horizontal index can be calculated as \cite{Emil2017}
\begin{align} \label{eq:horizontal-index}
 i(m) = (m-1) - M_{\rm h} \left\lfloor\frac{m-1}{M_{\rm  h}}\right\rfloor \in \{0,1,\ldots,M_{\rm h}-1\},
\end{align}
where $\lfloor \cdot \rfloor$ truncates the argument to the closest smaller integer.
The vertical index is similarly computed as
\begin{align} \label{eq:vertical-index}
j(m) = \left\lfloor\frac{m-1}{M_{\rm h}}\right\rfloor \in \{0,1,\ldots,M_{\rm  v}-1\}.
\end{align}
The system model above is illustrated in Fig.~\ref{figsystemmodel}. 

In a free-space LOS scenario with single-polarized antennas, the MIMO channel matrix $\mathbf{H}_{\rm u} \in \mathbb{C}^{M \times M}$ can be modeled as
\begin{equation} \label{eq:Hu}
\mathbf{H}_{\rm u}=    
\begin{bmatrix}
   \sqrt{\beta_{1,1}} e^{-\imaginary2\pi\frac{d_{1,1}-d}{\lambda}} & \cdots&\sqrt{\beta_{1,M}}e^{-\imaginary2\pi\frac{d_{1,M}-d}{\lambda}}\\
  \vdots &\ddots &\vdots \\
  \sqrt{\beta_{M,1}}e^{-\imaginary2\pi\frac{d_{M,1}-d}{\lambda}} & \cdots&\sqrt{\beta_{M,M}}e^{-\imaginary2\pi\frac{d_{M,M}-d}{\lambda}}\\
\end{bmatrix},
\end{equation}
where the phase-shifts are determined by the distance $d_{m,k}$ between transmit antenna $m$ and receive antenna $k$, the reference distance $d$, and the wavelength $\lambda$. The channel gain $\beta_{m,k}$ can be expressed as
\begin{equation}\label{beta_m_k}
\beta_{m,k} =  G^{\rm t}_{m,k} G^{\rm r}_{m,k} \left( \frac{ \lambda}{4 \pi d_{m,k}} \right)^2.
\end{equation}
This expression depends on the transmit and receive antenna gains, given by $ G^{\rm t}_{m,k}  $ and $  G^{\rm r}_{m,k}$, respectively. 
We can write the gains as $G^{\rm t}$ and $G^{\rm r}$ when they are constant over the whole array.
In the case of isotropic antennas, we have that $G^{\rm t}=G^{\rm r} =1$. The model in \eqref{eq:Hu} holds when each individual transmit antenna is in the far-field of each receive antenna, while the arrays can be in each other's radiative near-fields.

We will now extend the model in \eqref{eq:Hu} to consider two dual-polarized URAs. In the ideal case when the orthogonal polarizations are perfectly isolated, the corresponding free-space LOS channel matrix $\mathbf{H}_{\rm d} \in \mathbb{C}^{2M \times 2M}$ can be expressed as \cite{XSongthesis}
\begin{equation}\label{eq:H_d_def}
    \mathbf{H}_{\rm d}=\begin{bmatrix}
        \mathbf{H}_{\rm u} & \mathbf{0}\\
        \mathbf{0} & \mathbf{H}_{\rm u}
    \end{bmatrix}
    = \mathbf{I}_2 \otimes \mathbf{H}_{\rm u},
\end{equation}
where $\mathbf{I}_2$ is the $2 \times 2$ identity matrix and $\otimes$ denotes the Kronecker product. This matrix formulation is obtained when the antenna element indices $1,\ldots,M$ are associated with the first polarization while element indices $M+1,\ldots,2M$ belong to the second polarization. The zeros in $ \mathbf{H}_{\rm d}$ represent perfect isolation between the polarizations. 
Although the signal polarizations are maintained in free-space propagation, cross-talk generally appears in the transceiver hardware due to imperfect isolation \cite{Nabar2002a,coldrey2008modeling,Emil2017}. This is referred to as imperfect cross polar discrimination (XPD). We assume that each transmit antenna element radiates a fraction $(1-\gamma)$ of its power into the intended polarization and the remaining fraction $\gamma$ into the opposite polarization. The parameter $\gamma \in [0,1]$ characterizes the degree of imperfect XPD, where $\gamma=0$ is the ideal case.

We note that each antenna element captures a fraction $(1-\gamma)$ of the incident power of the signal with the intended polarization and a fraction $\gamma$ of the power from the opposite polarization. Consequently, when considering a pair of dual-polarized transmit and receive antennas, the fraction
\begin{equation} \label{eq:polarization1}
(1-\gamma)^2 + \gamma^2  = 1 - 2 (1-\gamma) \gamma 
\end{equation}
of the signal power reaches the receiver with the correct polarization \cite{LarssonP.2005,Zhou2012AADf}. By contrast, the fraction
\begin{equation} \label{eq:polarization2}
(1-\gamma) \gamma +  \gamma (1-\gamma)= 2 (1-\gamma) \gamma
\end{equation}
leaks into the opposite polarization, either at the transmitter or at the receiver.
Note that the sum of \eqref{eq:polarization1} and \eqref{eq:polarization2} equals $1$. Thus, the total signal power is maintained irrespective of the value of $\gamma$.
For brevity, we define 
\begin{equation}
\kappa \triangleq 2 (1-\gamma) \gamma
\end{equation}
to generalize the dual-polarized channel matrix in \eqref{eq:H_d_def} as 
\begin{align}\label{eq:H_d_def_kappa} \nonumber
    \mathbf{H}_{\rm d}&=\begin{bmatrix}
        \sqrt{1-\kappa}\mathbf{H}_{\rm u} & \sqrt{\kappa} \mathbf{H}_{\rm u}\\
        \sqrt{\kappa} \mathbf{H}_{\rm u} & \sqrt{1-\kappa}\mathbf{H}_{\rm u}
    \end{bmatrix} \\ &= \underbrace{\begin{bmatrix}
        \sqrt{1-\kappa} & \sqrt{\kappa}\\
        \sqrt{\kappa} & \sqrt{1-\kappa}
    \end{bmatrix}}_{\triangleq\mathbf{K}} \otimes \mathbf{H}_{\rm u},
\end{align}
 which is the MIMO channel that will be analyzed in the remainder of this paper. The matrix $\mathbf{K}$ characterizes the imperfect XPD.\footnote{The Kronecker product in (\ref{eq:H_d_def_kappa}) is obtained since we assume uniform XPD values across all antennas. If multiple antenna types are available to the array designer, it is logical to only use the one with the minimal XPD. However, in the hypothetical situation where antennas with varying XPD values are used, (\ref{eq:H_d_def_kappa}) does not hold and the  optimal spacing will likely be slightly different.}
Since the Frobenius norm of a Kronecker product is the product of the Frobenius norms, it follows that the channel matrix in \eqref{eq:H_d_def_kappa} has a squared norm of
\begin{equation}
\| \mathbf{H}_{\rm d} \|_{ \rm F}^2 = \| \mathbf{K} \|_{\rm  F}^2  \| \mathbf{H}_{\rm u} \|_{\rm  F}^2 = 2 \sum_{m=1}^{M} \sum_{k=1}^{M} \beta_{m,k}.
\end{equation}
This value is independent of the XPD in $\mathbf{K}$ and the phase-shifts of the individual elements in $\mathbf{H}_{\rm u}$ \cite{Liu2007}. Nevertheless, the MIMO channel capacity strongly depends on these parameters, because they determine how the value of $\| \mathbf{H}_{\rm d} \|_{\rm  F}^2$ is divided between the eigenvalues of $\mathbf{H}_{\rm d}^{\Htran}\mathbf{H}_{\rm d}$ \cite{Telatar1999a} (or equivalently, the singular values of $\mathbf{H}_{\rm d}$).
By using the conjugate transpose and distributive  properties of the Kronecker product, we obtain
\begin{equation}\label{Hdual}
    \mathbf{H}_{\rm d}^{\Htran}\mathbf{H}_{\rm d}= (\mathbf{K} \otimes \mathbf{H}_{\rm u})^{\Htran}(\mathbf{K} \otimes \mathbf{H}_{\rm u})=(\mathbf{K}^{\Htran}\mathbf{K}) \otimes (\mathbf{H}_{\rm u}^{\Htran}\mathbf{H}_{\rm u}).
\end{equation}
The elementwise products of the eigenvalues of $\mathbf{K}^{\Htran}\mathbf{K}$ and $\mathbf{H}_{\rm u}^{\Htran}\mathbf{H}_{\rm u}$ yield the eigenvalues of the matrix $\mathbf{H}_{\rm d}^{\Htran}\mathbf{H}_{\rm d} $. 

The eigenvalues of $\mathbf{K}^{\Htran}\mathbf{K} $ are
\begin{align} \label{eq:eigenvalues1}
\mu_1 &= 1+2\sqrt{(1-\kappa)\kappa}, \\
\mu_2 &= 1-2\sqrt{(1-\kappa)\kappa}, \label{eq:eigenvalues2}
\end{align}
which are obtained from its eigendecomposition: 
\begin{align}\nonumber
    \mathbf{K}^{\Htran}\mathbf{K} &= \begin{bmatrix}
        1 & 2\sqrt{\kappa}\sqrt{1-\kappa}\\
        2\sqrt{\kappa}\sqrt{1-\kappa} & 1
    \end{bmatrix} 
   \\&=
    \frac{1}{2} \begin{bmatrix}
        1 & -1\\
        1 & 1
    \end{bmatrix}
    \begin{bmatrix}
        \mu_1 & 0\\
        0 & \mu_2
    \end{bmatrix}
\begin{bmatrix}
        1 & 1\\
        -1 & 1
    \end{bmatrix}.\label{eq:15}
\end{align}
These eigenvalues depend on the XPD parameter $\kappa$, but they are independent of the antenna spacing.  Since this paper is not considering the design of dual-polarized antennas,  we will focus on optimizing the eigenvalues of $\mathbf{H}_{\rm u}^{\Htran}\mathbf{H}_{\rm u}$ to maximize the channel capacity by adjusting the antenna spacing.

\section{Capacity Maximization With Uniform Rectangular Arrays}\label{Sect_Capacity_max}

In this section, we will derive the antenna spacing for the URAs that optimizes the eigenvalues of the channel matrix to maximize the system capacity. First, we discuss URAs with different transmit and receive antenna spacing. Then, we consider a special case where both transmit and receive antenna arrays have the same spacing.  

\subsection{Optimal Antenna Spacing for URAs}

The URAs are deployed as illustrated in Fig.~\ref{figsystemmodel}. In particular, the antenna $m$ at the transmitter is located at the coordinate $(-i(m) {\rm h}_{\rm t}, -j(m) {\rm v}_{\rm t}, 0)$, while antenna $k$ at the receiver is located at the coordinate $(-i(k) {\rm h}_{\rm r}, -j(k) {\rm v}_{\rm r}, d)$.
Hence, the distance between these antennas can be obtained as
\begin{align}
d_{m,k}=\sqrt{d^2+\big(i(m){\rm h}_{\rm t}-i(k){ \rm h}_{\rm r}\big)^2+\big(j(m){\rm v}_{\rm t}-j(k){\rm v}_{\rm r}\big)^2},
 \end{align}
where we recall that ${{\rm v}_{\rm t}}$ and ${{\rm h}_{\rm t}}$ denote the vertical and horizontal spacing at the transmitter array, respectively. Similarly, the vertical and horizontal spacing at the receiver array are denoted as $  {{\rm v}_{\rm r}}$ and $  {{\rm h}_{\rm r}}$, respectively.
We notice that $d_{m,k}$ depends on all these parameters.
We denote the diagonal of the transmitter array as $D_{\rm t}$ and the diagonal of the receiver array as $D_{\rm r}$, expressed as
\begin{align}
\notag
   & D_{\rm t} = \sqrt{(M_{\rm h}-1)^2{\rm h}_{\rm t}^2+(M_{\rm v}-1)^2{\rm v}_{\rm t}^2},\\ 
    & D_{\rm r}=\sqrt{(M_{\rm h}-1)^2{\rm h}_{\rm r}^2+(M_{\rm v}-1)^2{\rm v}_{\rm r}^2},
\end{align}
respectively.
In typical propagation scenarios for which $d \geq 2D_{\rm  r}$ and $d \geq 2D_{\rm t}$, the channel gain is nearly the same between all antenna locations \cite{Bjornson2021a}:
\begin{equation}
    \beta_{m,k} \approx \beta = G^{\rm t}G^{\rm r} \left( \frac{\lambda}{4 \pi d} \right)^2. \label{eq:first-approx}
\end{equation}
By contrast, the phase-shift differences between the antennas cannot generally be neglected since these are relative to the wavelength. However, the exact expression can be simplified using the Fresnel approximation \cite{Bjornson2021a} (also known as the paraxial approximation \cite{Lozano2023}).
In particular, the distance between the transmit and receive antennas can be approximated by using the first-order Taylor series expansion as
\begin{align}\nonumber  
d_{m,k} &= d \sqrt{1+\frac{\big(i(m){\rm h}_{\rm t}-i(k){\rm h}_{\rm r}\big)^2 +\big(j(m){\rm v}_{\rm t}-j(k){\rm v}_{\rm r}\big)^2}{d^2}} 
\\&
\approx d+\frac{ \delta_{m,k} }{2d}, \label{eq:second-approx}
\end{align}
where $\delta_{m,k} = (  i(m){ \rm h}_{\rm t}-i(k){\rm h}_{\rm r} )^2 +( j(m){\rm v}_{\rm t}-j(k){\rm v}_{\rm r} )^2$. Qualitatively speaking, this approximation represents the utilization of a parabolic model instead of the precise spherical wavefront model \cite{parabolader}, but it is accurate for broadside communications.
By utilizing \eqref{eq:first-approx} and \eqref{eq:second-approx}, we obtain the Fresnel approximation of 
$\mathbf{H}_{\rm u}$ in \eqref{eq:Hu} as
\begin{equation} \label{eq:Hu_approx}
 \tilde{\mathbf{H}}_{\rm u} =  \sqrt{\beta}
\begin{bmatrix}
    e^{-\imaginary\pi\frac{\delta_{1,1}}{d\lambda}} & \cdots& e^{-\imaginary\pi\frac{\delta_{1,M}}{d\lambda}}\\
  \vdots &\ddots &\vdots \\
   e^{-\imaginary\pi\frac{\delta_{M,1}}{d\lambda}} & \cdots& e^{-\imaginary\pi\frac{\delta_{M,M}}{d\lambda}}\\
\end{bmatrix}.
\end{equation}
From \eqref{eq:Hu_approx}, it follows that $\| \Tilde{\mathbf{H}}_{\rm u} \|_{\rm  F}^2 = \beta M^2$ is the sum of the eigenvalues of $\Tilde{\mathbf{H}}_{\rm u}^{\Htran}\Tilde{\mathbf{H}}_{\rm u}$.

We will use the approximation in \eqref{eq:Hu_approx} for theory development and optimization, while the simulations in this paper will make use of the exact model in \eqref{eq:H_d_def_kappa}. In particular, we will use the approximation to find an antenna spacing that leads to $\Tilde{\mathbf{H}}_{\rm u}^{\Htran}\Tilde{\mathbf{H}}_{\rm u} = \beta \mathbf{I}_M$, so that all eigenvalues are equal and the high-SNR capacity is maximized.
The $ (l,k)$-th entry of the matrix  $\tilde{\mathbf{H}}_{\rm u}^{\Htran}\tilde{\mathbf{H}}_{\rm u}$ is given as
\begin{align}    
\beta \sum_{m=1}^{M} e^{\frac{\imaginary \pi }{d \lambda}(\delta_{m,l}-\delta_{m,k}) },
\label{eq:lkH}
\end{align}
where 
\begin{multline}    
    \delta_{m,k}= \big(  i(m){\rm  h}_{\rm t}-i(k){ \rm h}_{\rm r} \big)^2 +\big( j(m){\rm v}_{\rm t}-j(k){\rm v}_{\rm r} \big)^2
    \\
    = \bigr[(i(m){\rm  h}_{\rm t})^2 + (i(k){  \rm h}_{\rm r})^2 \bigl] +\bigr[(j(m){\rm  v}_{\rm t})^2 + (j(k){ \rm v}_{\rm r})^2 \bigl]\\
    -2\bigr[i(m)i(k){\rm h}_{\rm t}{ \rm h}_{\rm r}+j(m)j(k){\rm v}_{\rm t}{\rm v}_{\rm r} \bigl]. \label{deltamk}
\end{multline}
Therefore, we can write
\begin{multline}\label{eq_diff}
     \delta_{m,l}-\delta_{m,k}=
     \bigr[(i(l))^2-(i(k))^2 \bigl]{\rm h}_{\rm r}^2+\\
     \bigr[(j(l))^2-(j(k))^2 \bigl]{\rm v}_{\rm r}^2+2\bigr[(i(k)-i(l))i(m){\rm h}_{\rm t} {\rm h}_{\rm r}\\ +(j(k)-j(l))j(m){\rm v}_{\rm t}{\rm v}_{\rm r} \bigl]. 
\end{multline}
By substituting \eqref{eq_diff} into \eqref{eq:lkH}, we obtain
\begin{multline} \label{eq:offdiagonals_URA_Hu}
\beta \sum_{m=1}^{M} e^{\frac{\imaginary \pi }{d \lambda}(\delta_{m,l}-\delta_{m,k}) }=
\\A_{l,k} \sum_{m_{h}=1}^{M_{\rm  h}} e^{\frac{\imaginary 2\pi }{d \lambda}m_{\rm h}[i(k)-i(l)] { \rm h}_{\rm t}{\rm  h}_{\rm r}} \sum_{m_{\rm v}=1}^{M_{\rm v}} e^{\frac{\imaginary 2\pi }{d \lambda}m_{\rm v}[j(k)-j(l)] {\rm  v}_{\rm t}{\rm  v}_{\rm r}}, 
\end{multline} 
where $A_{l,k}= \beta e^{\frac{\imaginary \pi}{d \lambda} ([(i(l))^2-(i(k))^2 ]({\rm h}_{\rm r})^2 + [(j(l))^2-(j(k))^2 ]({\rm v}_{\rm r})^2) }$.
We recall that  $i(m)$ and $j(m)$ are the horizontal and vertical indices of antenna $m$, respectively. We notice that \eqref{eq:offdiagonals_URA_Hu} is separated into two distinct components: one associated with the horizontal antenna indices $i(k),i(l)$ and the other associated with the vertical antenna indices $j(k),j(l)$.

The diagonal entries of $\tilde{\mathbf{H}}_{\rm u}^{\Htran}\tilde{\mathbf{H}}_{\rm u}$ are equal to $\beta M$, as can be obtained from \eqref{eq:offdiagonals_URA_Hu} when $l=k$.
Since $|A_{l,k}|= \beta$, the off-diagonal entry (i.e., $l \neq k$) has a magnitude of
    \begin{align} \label{eq_off_diag} \nonumber
&\beta \left| \sum_{m_{\rm h}=1}^{M_{\rm h}} e^{\frac{\imaginary 2\pi }{d \lambda}m_{ \rm h}[i(k)-i(l)] {  h}_{t}{  h}_{r}} \right| \left| \sum_{m_{\rm v}=1}^{M_{\rm v}} e^{\frac{\imaginary 2\pi }{d \lambda}m_{\rm v}[j(k)-j(l)] {  v}_{t}{ v}_{r}} \right| \\ \nonumber
&=    
    \beta \left| \frac{1-e^{\imaginary\pi\frac{2M_{\rm h}(i(l)-i(k)) {\rm  h}_{\rm t}{ \rm h}_{\rm r}}{\lambda d}}}{1-e^{\imaginary\pi\frac{2(i(l)-i(k)) {\rm  h}_{\rm t}{\rm  h}_{\rm r}}{\lambda d}}} \right|\cdot \left| \frac{1-e^{\imaginary\pi\frac{2M_{\rm v}(j(l)-j(k)) { \rm v}_{\rm t}{ \rm v}_{\rm r}}{\lambda d}}}{1-e^{\imaginary\pi\frac{2(j(l)-j(k)) {\rm  v}_{\rm t}{\rm  v}_{\rm r}}{\lambda d}}} \right|,\\
&= \beta \cdot \left| \frac{\sin(\pi\frac{M_{\rm h} {\rm  h}_{\rm t}{ \rm h}_{\rm r}}{\lambda d})}{\sin(\pi\frac{{\rm  h}_{\rm t}{ \rm h}_{\rm r}}{\lambda d})}\right| \cdot \left|\frac{\sin(\pi\frac{M_{\rm v} {\rm  v}_{\rm t}{ \rm v}_{\rm r}}{\lambda d})}{\sin(\pi\frac{{\rm  v}_{\rm t}{ \rm v}_{\rm r}}{\lambda d})} \right|,
\end{align}
where the equality follows from using the classical geometric series formula in \cite{multi2011,Do2021a}. This is the product of two Dirichlet kernels, one depending on the horizontal antenna configuration (i.e., the number of antennas and spacings) and one depending on the vertical antenna configuration. This reveals that it is possible to decouple the optimization of the uniform rectangular array into distinct optimizations of horizontal and vertical uniform linear arrays, even if the horizontal and vertical DOFs are generally coupled \cite{Hu2018a,Pizzo2020a}.

\begin{Lemma} \label{lemma1}
    The  singular values of the MIMO channel matrix  $\tilde{\mathbf{H}}_{\rm u}$ are equal if the off-diagonal element of the matrix $\tilde{\mathbf{H}}_{\rm u}^{\Htran} \tilde{\mathbf{H}}_{\rm u}$ are zero. This happens if the antenna spacings satisfy
\begin{equation}\label{eq:8}
    { {\rm  h}_{\rm t}{\rm  h}_{\rm r}}=\frac{\lambda d}{M_{ \rm h}}, \quad \quad { { \rm v}_{\rm t}{  \rm v}_{\rm r}}=\frac{\lambda d}{M_{\rm  v}}.
\end{equation}
\end{Lemma}
\begin{IEEEproof}
    The off-diagonal element of the channel matrix in \eqref{eq_off_diag} becomes zero when either of the sine over sine ratios is zero. This happens when the denominator is non-zero and the numerator is zero, provided that both $M_{\rm h}$ and $M_{\rm v}$ are non-zero. In cases where either condition is not met, we can return to the first row in \eqref{eq_off_diag} to conclude that the expression becomes $\beta M$. To obtain the desired result regardless of the antenna indices, $\frac{M_{\rm h}  {\rm h}_{\rm t}  {\rm h}_{\rm r}}{\lambda d} = 1$ and $\frac{M_{\rm v}  {\rm v}_{\rm t}  {\rm v}_{\rm r}}{\lambda d} = 1$, must be satisfied simultaneously. Therefore, the spacing that yields the solution of \eqref{eq_off_diag}, referred to as the optimal spacing, can be expressed as shown in \eqref{eq:8}.
\end{IEEEproof}

This lemma is a variation on previous works that studied single-polarized arrays. Particularly, \cite{multi2011,Do2021a} considered ULAs (i.e., $M_{\rm v}=1$) with different spacings at the transmitter and receiver, while \cite{LarssonP.2005,Bohagen2007b,Zhou2012AADf} considered URAs.

Building on the fundamental result in Lemma~\ref{lemma1}, in the next theorem, we demonstrate that the optimal antenna spacing for the dual-polarized array is the same as for the single-polarized array.

\begin{Theorem} \label{theorem:capacity-maximization}
When the Fresnel approximation is tight, 
the high-SNR capacity with dual-polarized arrays is maximized using the antenna spacing in \eqref{eq:8}.
\end{Theorem}
\begin{IEEEproof}
    Under the Fresnel approximation, the channel matrix with dual-polarized antenna arrays in \eqref{eq:H_d_def_kappa} can be expressed as $\tilde{\mathbf{H}}_{\rm d} = \mathbf{K} \otimes \tilde{\mathbf{H}}_{\rm u}$. If we denote the singular values of $\mathbf{K}$ as $\sqrt{\mu_1},\sqrt{\mu_2}$ and the singular values of $\tilde{\mathbf{H}}_{\rm u}$ as $\sqrt{\lambda_1},\ldots,\sqrt{\lambda_{M}}$, then the $M$ singular values of $\tilde{\mathbf{H}}_{\rm d}$ are
    $\sqrt{\mu_i \lambda_m}$ for $i \in \{1,2 \}$ and $m \in \{1,\ldots,M\}$. At high SNRs, where equal power allocation is optimal, the capacity can be expressed as \cite{Tse2005a}
\begin{equation} \label{eq:capacityUpper1}
C_{\textrm{dual}}=\sum_{i=1}^{2}\sum_{m=1}^{M}\log_2 \left({1+\frac{P \mu_i \lambda_m }{ 2M \sigma^2}} \right),
\end{equation}
where $P$ denotes the total transmit power and $\sigma^2$ denotes the noise variance. By applying Jensen's inequality, we can rewrite \eqref{eq:capacityUpper1} as
\begin{align}       
C_{\textrm{dual}} &= \sum_{i=1}^{2} M \sum_{m=1}^{M} \frac{1}{ M} \log_2 \left({1+\frac{P \mu_i \lambda_m }{ 2M \sigma^2}} \right) \\
&\leq \sum_{i=1}^{2} M \log_2 \left({1+\frac{P \mu_i  }{ 2M \sigma^2} \frac{1}{ M} \sum_{m=1}^{M}\lambda_m} \right).
\end{align}
The upper bound is independent of how the singular values of $\tilde{\mathbf{H}}_{\rm d}$ are distributed since $\| \tilde{\mathbf{H}}_{\rm d} \|_{\rm  F}^2 = \sum_{m=1}^{M} \lambda_m$ and is achieved with equality if and only if all the singular values of $\tilde{\mathbf{H}}_{\rm d}$ are equal. This condition is satisfied when \eqref{eq:8} in Lemma~\ref{lemma1} holds. Hence, the optimal spacing with dual-polarized arrays is given by  \eqref{eq:8} and is independent of the singular values $\mu_1,\mu_2$ of the matrix $\mathbf{K}$. 
\end{IEEEproof}

This theorem proves that the optimal antenna spacing with dual-polarized arrays is the same as for single-polarized arrays, although we cannot make all the singular values equal in the dual-polarized case since the parameter $\kappa$ is constant. 
With the optimal spacing, we can write  $\tilde{\mathbf{H}}_{\rm u}^{\Htran}\tilde{\mathbf{H}}_{\rm u}=\beta M \mathbf{I}_M$. It follows that  $\tilde{\mathbf{H}}_{\rm d}^{\Htran}\tilde{\mathbf{H}}_{\rm d} = \mathbf{K}^{\Htran}\mathbf{K} \otimes \beta M \mathbf{I}_M$. 
Hence, the matrix $\tilde{\mathbf{H}}_{\rm d}^{\Htran}\tilde{\mathbf{H}}_{\rm d}$ has $M$ eigenvalues equal to $\mu_1\beta M$ and $M$ eigenvalues equal to $\mu_2 \beta M$, where $\mu_1,\mu_2$ are given in \eqref{eq:eigenvalues1}--\eqref{eq:eigenvalues2}.

\subsection{Geometric interpretation of the optimal spacing}

We will now interpret the results from Theorem~\ref{theorem:capacity-maximization} in light of near-field communications. 
Suppose the transmitter and receiver use dual-polarized arrays with the same form factor, such that  ${\rm  h}_{\rm t}= {\rm  h}_{\rm r}$ and ${\rm  v}_{\rm t}= {\rm  v}_{\rm r}$. Thus, it follows from \eqref{eq:8} that the optimal antenna spacing is
\begin{align} \label{eq:symmetric-separation}
    {\rm  h}_{\rm t}= {\rm  h}_{\rm r} =\sqrt{\frac{\lambda d}{M_{ \rm h}}}, \quad \quad
     { \rm v}_{\rm t} = {  \rm v}_{\rm r} =\sqrt{\frac{\lambda d}{M_{\rm  v}}}.
\end{align}
The spacing is much larger than the traditional $\lambda/2$-spacing, so the arrays are unusually sparse. Hence, we can build these arrays without being exposed to mutual coupling.

With the optimized antenna spacing, the arrays are typically operating in the radiative near-field, but not necessarily. The recent literature contains a multitude of ways to quantify this. In this context, the concept of finite-depth beamforming is most prevalent \cite{Ramezani2023b}. The Fraunhofer array distance $d_{\rm FA}$ for the system at hand is
\begin{equation}
    d_{\rm FA} =\frac{2 \left( (M_{ \rm h}{\rm  h}_{\rm r})^2+(M_{ \rm v}{\rm  v}_{\rm r} )^2 \right)}{\lambda} = 2d (M_{ \rm h} + M_{ \rm v}).
\end{equation}
Beams are known to have a finite depth when they are focused on a distance that satisfies $d \leq d_{\rm FA} /10$. This condition holds whenever $M_{ \rm h} + M_{ \rm v} \geq 5$, which is the case in most of the anticipated use cases for the provided theory. However, it is not the finite beamdepth or spherical wavefronts that enables a high-rank channel matrix, but the tiny beamwidth.

\begin{figure}[t!]
			\begin{overpic}[width=\columnwidth,tics=10]{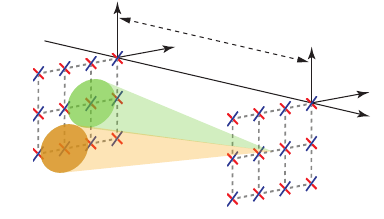}
                \put(28,15){\small \rotatebox{5}{Beam on (3,3)}}
				\put(35,28){\small \rotatebox{-20}{Beam on (2,2)}}
    		    \put(55,48.5){\small $d$}
    		    \put(8,5){\small Transmitter}
    		    \put(76,2){\small Receiver}
			\end{overpic}  
\caption{With the optimized antenna spacing from Theorem~\ref{theorem:capacity-maximization}, the receiver can focus beams on specific transmit antennas while creating nulls at all other antennas. This is sketched by showing two main lobes focused on two different antennas.}
\label{beamwidth_illustration}
\end{figure}
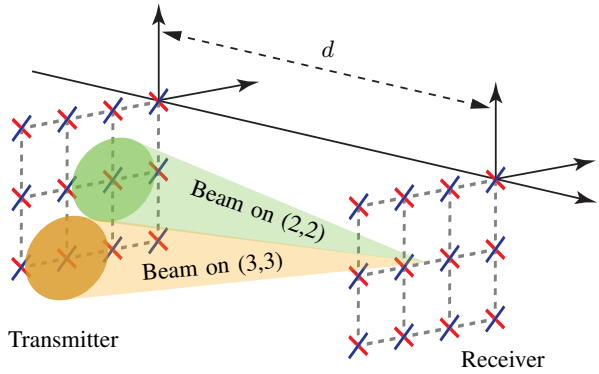

The spacing in \eqref{eq:symmetric-separation} grows with the wavelength $\lambda$ and propagation distance $d$ but decays with the total number of antennas. There are physical reasons for this.
The horizontal first-null beamwidth of a horizontal ULA with $M_{ \rm h}$ elements is \cite[Sec.~7.4]{Emil2017}
\begin{equation}
\mathrm{BW} = 2 \arcsin \left( \frac{\lambda}{M_{ \rm h} {\rm  h}_{\rm r}} \right) = 2 \arcsin \left(  \sqrt{\frac{\lambda}{M_{ \rm h} d}} \right).
\end{equation}
Hence, when the receiver processing focuses a beam on a transmit antenna that is a distance $d$ away, the beam has a physical width of
\begin{equation}
    d \cdot \tan(\mathrm{BW})  \approx 2 d   \sqrt{\frac{\lambda}{M_{ \rm h} d}} = 2 \sqrt{\frac{\lambda d}{M_{ \rm h}}} = 2 {\rm  h}_{\rm t},
\end{equation}
where the approximation utilizes the facts that $\tan(\cdot)$ and $\arcsin(\cdot)$ are linear function when the argument is small. Since the physical beamwidth is $2 {\rm  h}_{\rm t}$, there will be nulls at the adjacent antennas. In fact, with the optimized antenna spacing, when the receiver focuses a beam at one antenna, there will be nulls for signals arriving for all other transmit antennas. This is why the capacity is maximized: we get the maximum beamforming gain without interference.
When this principle is applied in two dimensions, using URAs, we get the behavior sketched in Fig.~\ref{beamwidth_illustration} where each ``beam'' has a main lobe focused on amplifying signals received from one specific antenna direction without covering any other antennas. There will be side-lobes outside the main lobe, but the nulls appear at the other antenna locations.
We need the combination of a transmitter with a large antenna spacing and a receiver that creates small beams (by also having a large spacing).

When the antenna spacing is larger than half-wavelength, the main lobe becomes narrower than in the $\lambda/2$ case, but the beamforming remains the same. The removed parts of the main lobe appear at very different angles and are called grating lobes. While conventional side lobes are much weaker than the main lobe, grating lobes achieve the full beamforming gain. This can cause interference to other systems (e.g., in multi-user setups) but does not lead to power losses for the point-to-point channel because $\| \mathbf{H} \|_F^2$ remains the same regardless of the antenna spacing. Moreover, despite the tiny beamwidth, the system will be as sensitive to interference from other sources as a physically smaller array with the same number of antennas.

\subsection{Capacity with optimized dual-polarized URAs}
\label{subsec:optimal_capacity}

For a given maximum transmit power $P$ and noise variance $\sigma^2$, the channel capacity with the optimal antenna spacing is 
\begin{equation}\label{eq:24}          
C_{\textrm{dual}}=\sum_{i=1}^{2}\sum_{m=1}^{M}\log_2 \left({1+\frac{q_{ m,i} \mu_{i} \beta M }{\sigma^2}} \right),
\end{equation}
where the power allocation $q_{1,i},\ldots,q_{M,i}$ for $i=1,2$ is selected using the water-filling algorithm \cite{Telatar1999a}. Since there are two eigenvalues with multiplicity $M$, the water-filling power allocation becomes 
\begin{align} \label{eq:waterfilling1}
q_{m,1} &= \begin{cases} 
      \frac{P}{M}, &  P \le \frac{\sigma^2}{\mu_{2} \beta}-\frac{\sigma^2}{\mu_{1} \beta}, \\  
      \frac{P}{2M}+\frac{\sigma^2}{2\mu_{2} \beta M}-\frac{\sigma^2}{2\mu_{1} \beta M} & \textrm{otherwise},
   \end{cases} \\ \label{eq:waterfilling2}
q_{m,2} &= \begin{cases} 
      0, &  P \le \frac{\sigma^2}{\mu_{2} \beta}-\frac{\sigma^2}{\mu_{1} \beta}, \\  
      \frac{P}{2M}+\frac{\sigma^2}{2\mu_{1} \beta M}-\frac{\sigma^2}{2\mu_{2} \beta M} & \textrm{otherwise}. 
   \end{cases}
\end{align}
At low SNRs, the capacity is achieved by dividing the power equally over the $M$ instances of the strongest eigenvalue.
In the medium- and high-SNR regimes, which are jointly represented by $P > \frac{\sigma^2}{\mu_{2} \beta}-\frac{\sigma^2}{\mu_{1} \beta}$, all the transmit power values are non-zero. The capacity expression in \eqref{eq:24} then becomes
\begin{multline}
    C_{\textrm{dual}} = M \log_2 \left(1+\frac{P \mu_{1} \beta  }{2\sigma^2}+\frac{ \mu_{1}-\mu_{2}}{2\mu_{2}}  \right)\\ + M \log_2 \left(1+\frac{P \mu_{2} \beta  }{2\sigma^2}+\frac{ \mu_{2}-\mu_{1}}{2\mu_{1}}  \right). \label{eq:C-expression}
\end{multline}
This capacity is exactly proportional to the number of dual-polarized antennas, $M$. This happens despite the fact that the total transmit power $P$ is divided over $2M$ eigendirections, because the receive beamforming gain also increases linearly with $M$ so that the received power per eigendirection is constant.
The multiplexing gain is $2M$ since there are that many logarithmic terms, even if half of them are smaller than the other half.

In the special case of perfect XPD (i.e., $\kappa=0$), we have $\mu_1 = \mu_2 = 1$ and the capacity expression in \eqref{eq:C-expression} simplifies to $2M \log_2 (1+\frac{P  \beta  }{2\sigma^2} )$. The multiplexing gain is still $2M$ and the transmit power is equally divided between the two polarizations. 
The channel matrix with the optimal spacing satisfies
$\mathbf{H}_{\rm d}^{\Htran}\mathbf{H}_{\rm d} = \beta M \mathbf{I}_{2M}$, which implies that the capacity is achieved by transmitting an independent signal from each antenna and from each polarization dimension.

In practice, we encounter a different case where $\kappa > 0$.
When $\kappa$ increases towards $0.5$, the capacity is monotonically decreasing due to the imbalanced eigenvalues in \eqref{eq:eigenvalues1}--\eqref{eq:eigenvalues2} of the matrix $\mathbf{K}^{\Htran}\mathbf{K}$. We can establish the following result by utilizing Jensen's inequality as in the proof of Theorem~\ref{theorem:capacity-maximization}.

\begin{corollary} \label{cor:kappa}
The high-SNR capacity is maximized for $\kappa \in [0,1]$ when $\kappa=0$ or $\kappa=1$, leading to $\mu_1=\mu_2=1$.
\end{corollary}

This corollary proves that it is optimal to have perfect XPD, which might seem obvious but it is not a trivial result. At low SNRs, where the water-filling power allocation only allocates power to the strongest eigendirections of the channel, it is instead optimal to have $\kappa=0.5$ so that $\mu_1=2$ and $\mu_2=0$, because \eqref{eq:waterfilling1} implies that we only assign power to the $M$ strongest eigenvalues.

\begin{figure}[t!]
\includegraphics[width=\textwidth]{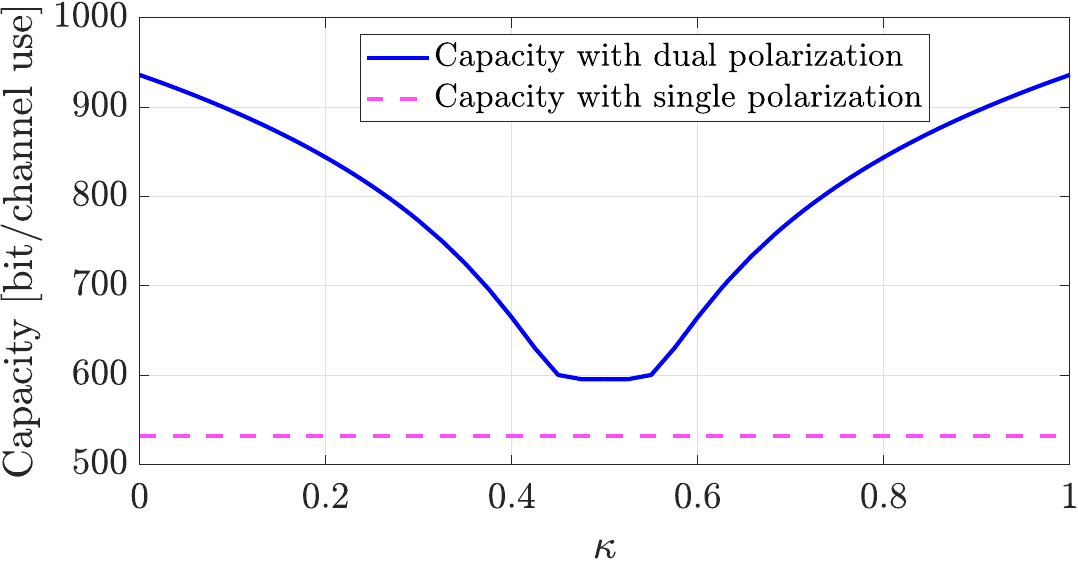}
\caption{The channel capacity vs. the XPD-parameter $\kappa$ when using $64$ dual-polarized antennas, for which the optimal antenna spacing is $0.35$\,m, and the transmitter/receiver array has the area $6.18$\,m$^2$. The capacity with $128$ single-polarized antennas is shown as a reference.}
\label{figkappa}
\end{figure}

Fig.~\ref{figkappa} shows the high-SNR channel capacity as a function of $\kappa$ when using $M=64$ dual-polarized antennas. We consider an SNR of $P\beta/\sigma^2 =25$ dB, propagation distance of $d=100$\,m, carrier frequency of $30$\,GHz, and equal square-shaped arrays at the transmitter and receiver. 
The results confirm that a dual-polarized system achieves nearly twice the capacity of the corresponding single-polarized system (with perfect XPD and $M=128$ single-polarized antennas) when $\kappa$ is close to $0$. The same happens in the hypothetical case of $\kappa \approx 1$ when the two polarizations are well separated at the receiver but swapped. There is a gain of having dual-polarized antennas even when $\kappa \approx 0.5$ thanks to the increased beamforming gain.

The Kronecker structure $\tilde{\mathbf{H}}_{\rm d}^{\Htran} \tilde{\mathbf{H}}_{\rm d} = \mathbf{K}^{\Htran}\mathbf{K} \otimes \beta M \mathbf{I}_M$ along with the eigendecomposition in \eqref{eq:15} implies that independent signals should be transmitted from each of the $M$ antenna locations. However, the co-located dual-polarized elements do not transmit independent signals to achieve capacity. Instead, the strongest eigenvalue $\mu_1 M$ is achieved by transmitting the same signal identically from both polarizations, while the weaker eigenvalue $\mu_2 M$ is achieved by transmitting the same signal but with opposite signs using the two polarizations.
These ways of transmitting are respectively given by the eigenvectors $[1,1]^{\Ttran}/\sqrt{2}$ and $[-1,1]^{\Ttran}\sqrt{2}$ in \eqref{eq:15}.

The XPD is often quantified as $(1-\kappa)/\kappa$ and typical measured values ranges from 5 to 15\,dB, with the largest values for outdoor LOS-like channels and the smallest values for rich-scattering NLOS channels (indoor or outdoor) \cite{coldrey2008modeling}.

\begin{figure}[t!]
\centering
\subfloat[Channel capacity vs. antenna spacing.]
{\includegraphics[width=\textwidth]{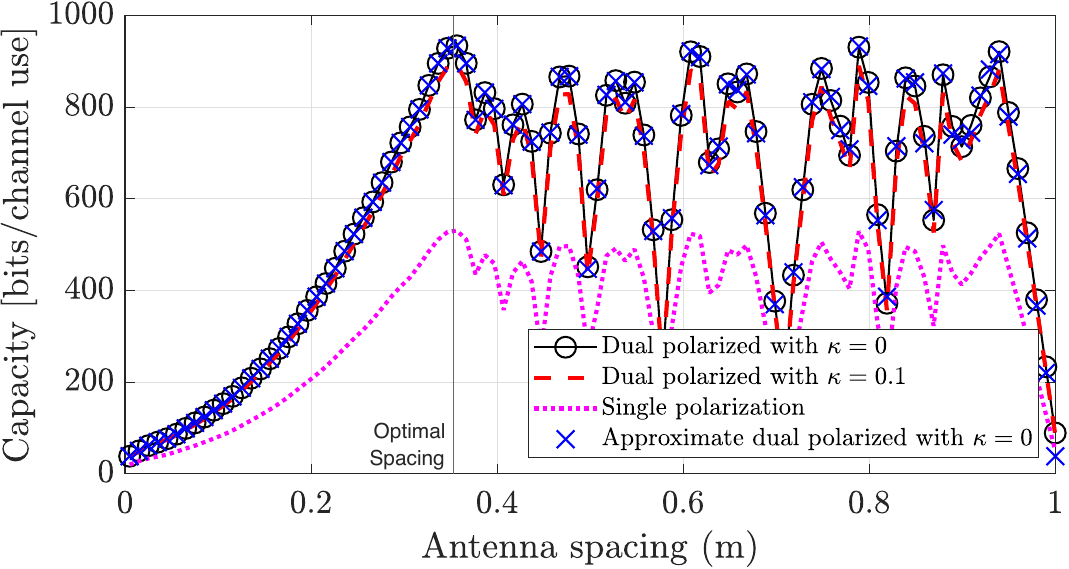}}\hfill
\centering
\subfloat[Rank vs. antenna spacing]
{\includegraphics[width=\textwidth]{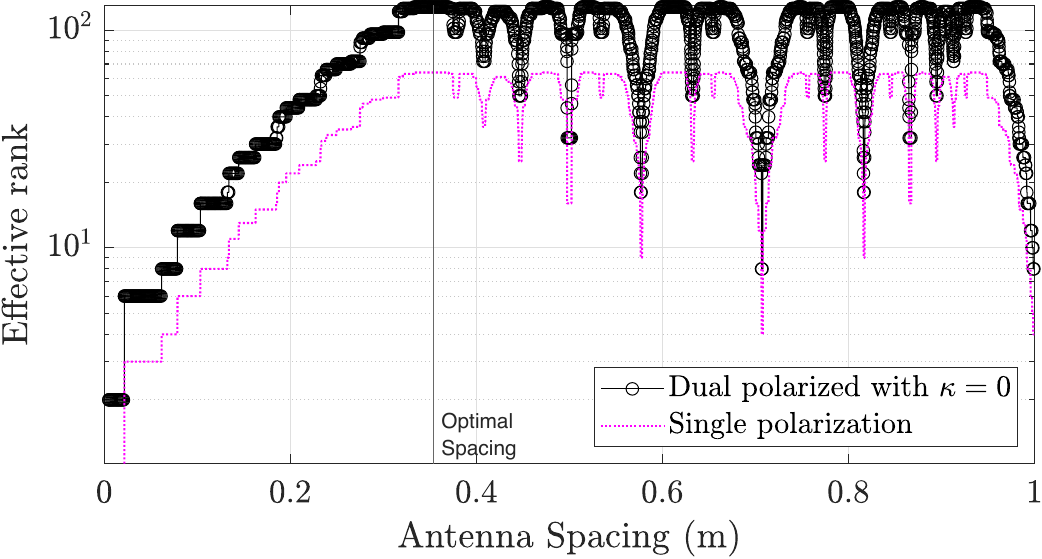}}\hfill
\centering
\subfloat[Condition number vs. antenna spacing]
{\includegraphics[width=\textwidth]{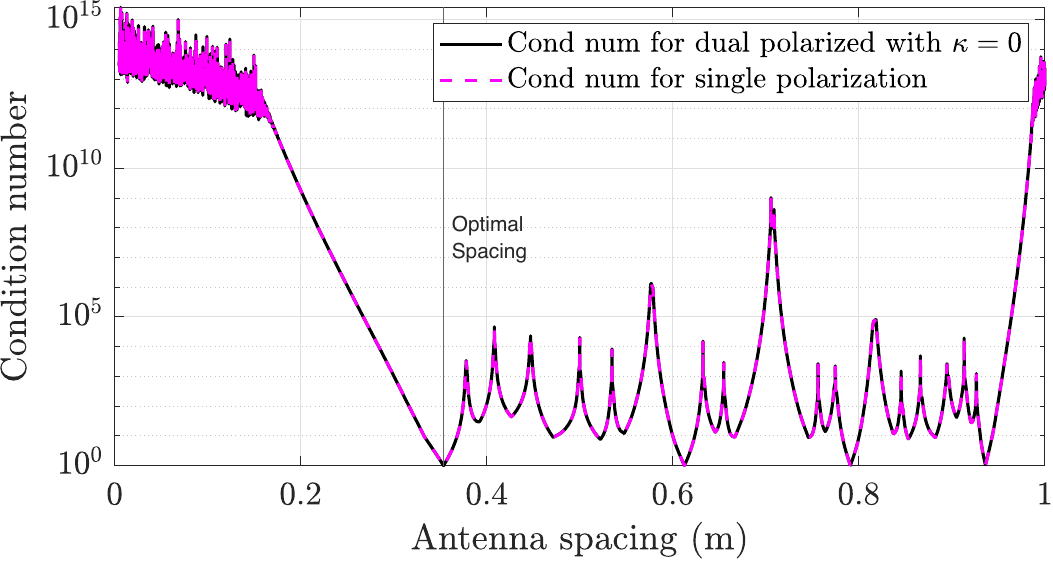}}
\caption{The channel capacity as a function of the horizontal/vertical antenna spacing for a URA with $M_{\rm h}=M_{\rm v}=8$ antennas and equal spacing at the transmitter and receiver where ${A}_{\rm t}={A}_{\rm r}$ varies from 0.002 to 49 $m^2$ .}
\label{fig4}
\vspace{-4mm}
\end{figure}

The relation between the antenna spacing and MIMO channel capacity is demonstrated in Fig.~\ref{fig4}(a) for the same setup as in Fig.~\ref{figkappa}.
We consider square-shaped arrays with the common antenna spacing $\Delta = {\rm h}_{\rm t}={\rm h}_{\rm r}={\rm v}_{\rm t}={\rm v}_{\rm r}$, which is shown on the horizontal axis.
We notice that increasing the antenna spacing first monotonically improves the MIMO capacity until we reach the maximum capacity of $\approx 900$ bit/channel use at $\Delta = 0.35$\,m, which is the optimal spacing from \eqref{eq:symmetric-separation}. This is a seemingly extraordinary capacity value but it 
is achieved using $128$ orthogonal spatial dimensions that each carry $\approx7$ bit/dimension and can be implemented using coded 256-QAM, which is practically feasible.
When the antenna spacing is further increased, the capacity fluctuates with some peaks equally good as at the optimal point.
This highlights the importance of using the optimal spacing, while the price to pay is that each array is $2.5 \times 2.5$ meters large at the optimal point in this mmWave setup with a range of 100 meters. Since the array dimensions are proportional to $\sqrt{\lambda d}$, we can shrink the arrays by reducing the propagation distance or wavelength. The XPD leads to a minor capacity reduction, as can be seen by comparing the curves with $\kappa=0$ and $\kappa=0.1$. As expected, the maximum capacity appears at the same antenna spacing regardless of $\kappa$. The notches in Fig.~\ref{fig4} do not drop to the same level, which is not due to a lack of simulation resolution, but a consequence of the effective channel rank dropping to different values (all larger than one).

The results in Fig.~\ref{fig4} are obtained using the exact channel matrix model in \eqref{eq:H_d_def}, but the solid line is obtained using the Fresnel approximation in \eqref{eq:Hu_approx} that was used to obtain the analytical results. The accuracy of the approximation is readily visible, with the approximate curve closely matching the actual curve.
Finally, we notice that the capacity is nearly doubled when using dual polarization compared to using a single polarization. 

Fig.~\ref{fig4}(b) shows how the effective rank (defined as the sum of singular values greater than the threshold $\epsilon=1$) varies with the antenna spacing, while 
Fig.~\ref{fig4}(c) shows how the condition number varies. 
We observe that the maximum capacity is achieved at the optimal antenna spacing characterized by the minimum condition number and full effective rank. When the capacity is reduced, it is first the condition number that increases and then the effective rank decreases.
When the rank is full but the capacity is less than maximum, the singular values are unequal because of grating lobes; the signals transmitted from some antennas are partially mixed with the signals transmitted from other antennas so we cannot make the perfect receiver separation that was illustrated in Fig.~\ref{beamwidth_illustration}. It is only at the optimal antenna spacing or less that the communication channel is entirely unaffected by the grating lobes.

\section{Array Geometry Optimization}

The capacity expressions in \eqref{eq:24} and \eqref{eq:C-expression} depend on the total number of antenna elements $M= M_{\rm h} M_{\rm v}$, regardless of how these are divided between the rows and columns in the URA. Therefore, we have the degrees of freedom to determine the values of $M_{\rm h}$ and $M_{\rm v}$ that satisfy $M= M_{\rm h} M_{\rm v}$ while optimizing the array geometries to enable easier deployment.

In this section, we will consider two different optimization metrics: 1) minimal total aperture length; and 2) minimal total aperture area. Having a small aperture length (i.e., diagonal of the array) is important when the deployment location has a specific maximal dimension that must be complied with, but we can rotate the array arbitrarily to do that. On the other hand, the wind load on an array is determined by the aperture area and can be the limiting factor for base station deployment.

We want to optimize the array geometries while adhering to the optimal antenna spacing condition in \eqref{eq:8}. Without loss of generality, we introduce the variables $\alpha,\gamma \in [0,1]$, to parametrize all possible optimal antenna spacings as follows:
\begin{align}\label{eq_decompose1}
&{ {\rm  h}_{\rm t}{\rm  h}_{\rm r}}=\frac{\lambda d}{M_{ \rm h}} = \underbrace{\left( \frac{\lambda d}{M_{\rm  h}}\right)^\alpha}_{{\rm  h}_{\rm t}}  \underbrace{\left( \frac{\lambda d}{M_{\rm  h}}\right)^{(1-\alpha)}}_{{\rm  h}_{\rm r}},
\\
& { {\rm  v}_{\rm t}{\rm  v}_{\rm r}}=\frac{\lambda d}{M_{ \rm v}} = \underbrace{\left( \frac{\lambda d}{M_{\rm  v}}\right)^\gamma}_{{\rm  v}_{\rm t} }   \underbrace{ \left( \frac{\lambda d}{M_{\rm  v}}\right)^{(1-\gamma)}}_{{\rm  v}_{\rm r}}. \label{eq_decompose2}
\end{align}
We define the horizontal length $L^{\rm h}_{\rm t}$ and vertical length $L^{\rm v}_{\rm t}$ of the transmitter's URA are calculated using \eqref{eq:8} as
\begin{align}\label{eq:20}
    L^{\rm h}_{\rm t}&={ \rm h}_{ \rm t}(M_{\rm  h}-1)+W = \frac{\lambda d}{M_{ \rm h}{\rm  h}_{ \rm r}}(M_{\rm  h}-1)+W, \\
    L^{\rm v}_{\rm t}&={\rm   v}_{\rm  t}(M_{ \rm v}-1)+W = \frac{\lambda d}{M_{\rm  v}{ \rm v}_{\rm  r}}(M_{ \rm v}-1)+W,
\label{eq:21}
\end{align}
where $W>0$ is the width of an individual antenna element. Similarly, we express the horizontal length $L^{\rm h}_{\rm r}$ and vertical length $L^{\rm v}_{\rm r}$  of the receiver's URA as 
\begin{align}\label{eq:20b}
    L^{\rm h}_{\rm r}&={\rm  h}_{ \rm r}(M_{\rm  h}-1)+W = \frac{\lambda d}{M_{ \rm h}{ \rm  h}_{\rm t}}(M_{\rm  h}-1)+W, \\
    L^{\rm v}_{\rm r}&={ \rm v}_{\rm  r}(M_{\rm  v}-1)+W = \frac{\lambda d}{M_{\rm  v}{ \rm  v}_{\rm t}}(M_{\rm  v}-1)+W.
\label{eq:21b}
\end{align}

\begin{figure*}
\centering
\subfloat[Aperture length versus the number of antenna elements.] 
{\includegraphics[width=0.49\textwidth]{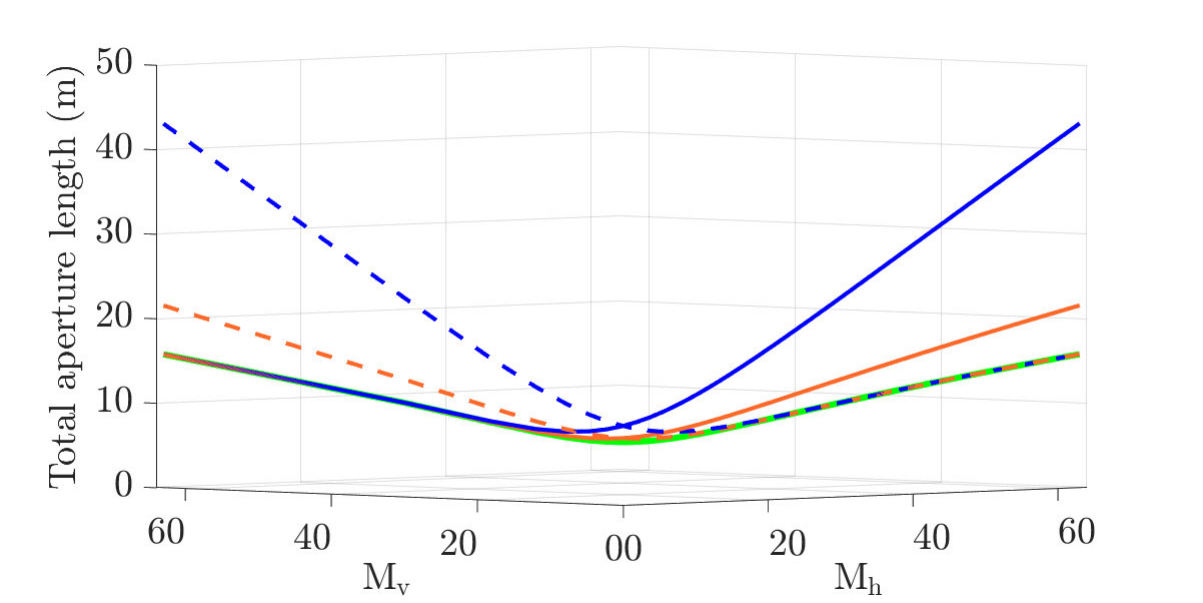}\label{aperturelength}}\hfill
\subfloat[Aperture area versus the number of antenna elements.]
{\includegraphics[width=0.49\textwidth]{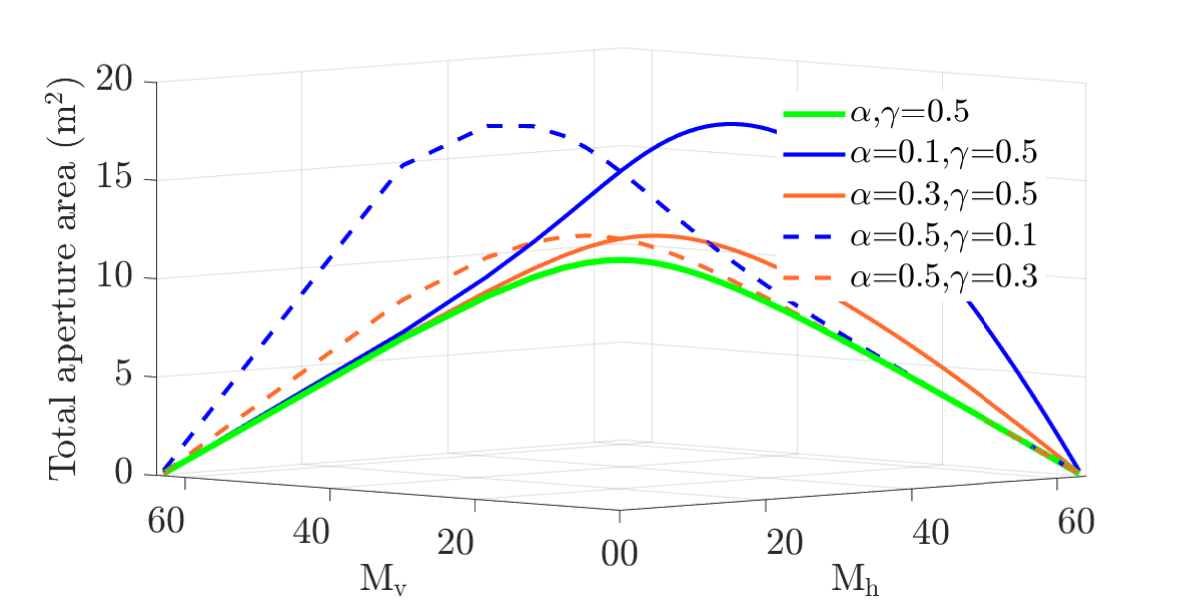}\label{aperturearea}}\hfill
\centering
\caption{The aperture length and aperture area for different values of $M_{\rm h}$, $M_{\rm v}$, $\alpha$ and $\gamma$.}
\label{Aperturelengthandarea}
\end{figure*}

\subsubsection{Aperture Length Minimization}

Our objective is now to minimize the total aperture length at the transmitter and receiver, expressed as
\begin{equation}
    \sqrt{(L^{\rm h}_{\rm t})^2+(L^{\rm v}_{\rm t})^2}+\sqrt{(L^{\rm h}_{\rm r})^2+(L^{\rm v}_{\rm r})^2}.
\end{equation}
By using the parameters $\alpha$ and $\gamma$, we can define a function for the total aperture length as follows:
\begin{align}
\notag
    &f(\alpha, \gamma,M_{\rm h},M_{\rm v})= \\   \notag&\sqrt{\left({\rm h_t}(M_{\rm h}-1)+W\right)^2+ 
    \left({\rm v_t}(M_{\rm v}-1)+W \right)^2}+  
     \\
    &\sqrt{ \left(  {\rm h_r}(M_{\rm h}-1)+W\right)^2+
    \left({\rm v_r}(M_{\rm v}-1)+W \right)^2}. \label{aperturecalc}
\end{align}
The corresponding optimization problem can be expressed as 
\begin{equation}\label{eq:aperture-problem}
\begin{aligned}
& \underset{\alpha,\gamma, M_{ \rm v},M_{\rm  h}  }{\text{minimize}}
& &f(\alpha,\gamma,M_{\rm h},M_{\rm v})\\
& \, \text{subject to}
& & M = M_{\rm  h} M_{\rm  v}, 0\leq\alpha\leq 1,0\leq\gamma\leq 1.
\end{aligned}
\end{equation}
To solve this problem, we will consider the variables sequentially. We begin by considering $\alpha$ and $\gamma$ and taking the derivative of $f(\alpha,\gamma, M_{ \rm v},M_{\rm  h})$ with respect to $\gamma$. The derivative is given in \eqref{aperturederivative} at the bottom of the next page.

{

\begin{figure*}[b!]
\hrule
\smallskip
\begin{multline}
    \frac{d}{d\gamma}f(\alpha,\gamma,M_{\rm h},M_{\rm v})= 
    \frac{\log(\frac{\lambda d}{M_{\rm v}})\left( {\rm v_t}(M_{\rm v}-1)+W\right)(M_{\rm v}-1){\rm v_t}}{\sqrt{\left( {\rm h_t}(M_{\rm h}-1)+W\right)^2+ \left( {\rm v_t} (M_{\rm v}-1)+W \right)^2}}
    -\frac{\log(\frac{\lambda d}{M_{\rm v}})\left( {\rm v_r}(M_{\rm v}-1)+W\right)(M_{\rm v}-1){\rm v_r} }{\sqrt{\left(  {\rm h_r}(M_{\rm h}-1)+W\right)^2+ \left({\rm v_r} (M_{\rm v}-1)+W \right)^2}}
    \label{aperturederivative}
\end{multline}
\end{figure*}
}

The derivative is zero if $\gamma=\alpha=0.5$, which can be shown to be a minimum. If $\alpha \neq 0.5$, then the derivative will be zero at another $\gamma$-value. However, since the derivative with respect to $\alpha$ has the shape in \eqref{aperturederivative}, stated on the text page, the symmetry implies that $\gamma=\alpha=0.5$ is the solution. This means that to minimize the total aperture length, the two arrays should have identical sizes. The intuition is that if we shorten one array, the length of the array will increase disproportionately due to the multiplication in 
\eqref{eq_decompose1}--\eqref{eq_decompose2}.

If we substitute $\alpha=\gamma=0.5$ into \eqref{aperturecalc}, we obtain
\begin{align} \nonumber
&f(M_{\rm h},M_{\rm v})=
\\
& 2\sqrt{\left( \sqrt{ \frac{\lambda d}{M_{\rm  h}}}(M_{\rm h}-1)+W\right)^2+ \left( \sqrt{\frac{\lambda d}{M_{\rm  v}}}(M_{\rm v}-1)+W \right)^2}. \label{func_value_mhmv}
\end{align}
Since $M_{\rm v} = M/M_{\rm h}$ is needed to satisfy the constraint, we can substitute this into \eqref{func_value_mhmv} to eliminate $M_{\rm v}$. The derivative with respect to $M_{\rm h}$ can be written as in \eqref{funcwrtMh}, and  it equals zero when $M_{\rm  h}=\sqrt{M}$. This can be shown to be a minimum. Consequently, the solution to the optimization problem \eqref{eq:aperture-problem} is $M_{\rm  v}= M/M_{\rm h}=\sqrt{M}$ and $\alpha=\gamma=0.5$.

\begin{figure*}

\begin{align}
      \frac{d}{dM_{\rm h}}f(M_{\rm h})=
    \frac{-\frac{\lambda d}{M_{\rm  h}^2}+\frac{W \sqrt{d \lambda M_{\rm  h}}}{M_{\rm  h}}+\frac{W \sqrt{\frac{d \lambda}{M_{\rm  h}}}}{M_{\rm  h}}+d\lambda-\frac{\lambda d M}{M_{\rm  h}^2}-\frac{W \sqrt{\frac{d \lambda M_{\rm  h}}{M}}}{M_{\rm  h}}-\frac{W \sqrt{\frac{d \lambda M}{M_{\rm  h}}}}{M_{\rm  h}}+\frac{d\lambda}{M}}{\sqrt{\left(\sqrt{  \frac{\lambda d}{M_{\rm  h}}}(M_{\rm h}-1)+W\right)^2+ \left(\sqrt{ \frac{\lambda d M_{\rm  h}}{M}}(\frac{M}{M_{\rm h}}-1)+W \right)^2}}
    \label{funcwrtMh}
\end{align}
\hrulefill
\end{figure*}

In summary, the total aperture length is minimized by having uniform square arrays (USAs) of equal size at the transmitter and receiver.
We illustrate this result in Fig.~\ref{aperturelength}, where the total aperture length is shown as a function of $M_{\rm h}$ and $M_{\rm v}$ with $M=64$ antenna elements.
The propagation distance is $d=100$\,m. The width of each antenna element is  $W=\lambda/2$ and the carrier frequency is $30$ GHz. 
The figure shows that the total length is maximized by ULAs. On each curve, the minimum is in the middle, where the arrays are square-like. The global minimum is obtained when $\alpha=\gamma=0.5$ in the case of USAs with $M_{\rm h}=M_{\rm v}$.

\subsubsection{Aperture Area Minimization}

The array geometries can also be designed to minimize the total area, which leads to a different solution.
To proceed, we first notice that the total area of the transmitter and receiver arrays can be expressed as 
 \begin{align}\nonumber
     A_{\rm t}+A_{\rm r} &= L^{\rm h}_{\rm t} L^{\rm v}_{\rm t} + L^{\rm h}_{\rm r} L^{\rm v}_{\rm r} \\ \nonumber
     &=\left( {\rm h}_{\rm t}(M_{\rm h}-1)+W\right) \left( {\rm v}_{\rm t}(M_{\rm v}-1)+W \right)
     \\&+\left( {\rm h}_{\rm r}(M_{\rm h}-1)+W\right) \left( {\rm v}_{\rm r}(M_{\rm v}-1)+W \right).
     \label{area}
 \end{align}
The area minimization problem can be formulated as
\begin{equation}\label{eq:totalareamini}
\begin{aligned}
& \underset{\alpha,\gamma,M_{\rm  v},M_{\rm  h} \in \{1,\ldots,M\}}{\text{minimize}}
& & {A}_{\rm t}+{A}_{\rm r} \\
& \,\,\,\,\,\,\,\,\,\,\, \text{subject to}
& & M = M_{ \rm\rm h} M_{ \rm v}, 0\leq\alpha\leq 1,0\leq\gamma\leq 1,
\end{aligned}
\end{equation}
where 
\begin{align}\nonumber
    A_{\rm t}+A_{\rm r}=\left({\rm h_t}(M_{\rm h}-1)+W\right) \left({\rm v_t}(M_{\rm v}-1)+W\right)
    \\+\left({\rm h_r}(M_{\rm h}-1)+W\right)\left( {\rm v_r}(M_{\rm v}-1)+W\right) \label{areaderivative}
\end{align}
follows from substituting \eqref{eq_decompose1}--\eqref{eq_decompose2} into \eqref{area}.
The first derivative of \eqref{areaderivative} with respect to $\gamma$ is given in \eqref{areade} at the top of the next page.

\begin{figure*}
\begin{align}
    \frac{d}{d\gamma}(A_{\rm t}+A_{\rm r})= \left(\left({\rm h_t}(M_{\rm h}-1)+W\right)(M_{\rm v}-1){\rm v_t}\log\left(\frac{\lambda d}{M_{\rm v}}\right)\right)
    -\left(\left({\rm h_r}(M_{\rm h}-1)+W\right)(M_{\rm v}-1){\rm v_r}\log\left(\frac{\lambda d}{M_{\rm v}}\right)\right) 
   \label{areade}
\end{align}
\hrulefill
\end{figure*}

The derivative in \eqref{areade} equals zero when both $\alpha$ and $\gamma$ are $0.5$. Further solutions can be found where the two parameters are different, but this is the minimum since the symmetry with respect to these parameters requires them to be equal.
We notice that the values of  $\alpha$ and $\gamma$ minimizing the total aperture area are the same as those minimizing the total aperture length; that is, the transmitter and receiver should have equal-shaped arrays. 

By substituting $\alpha=\gamma=0.5$ and $M_{\rm v} = M/M_{\rm h}$ into \eqref{areaderivative}, we can obtain the area expression
\begin{equation}\label{new_areaderivative}
\frac{ W \sqrt{ \lambda d M_{\rm h}} \left(1 + M_{\rm h} -\sqrt{M} - \frac{M_{\rm h}}{\sqrt{M}} \right) + 2 \frac{\lambda d}{M} \left(M - M_{\rm h}^2 \right)}{ M_{\rm h}^2},
\end{equation}
which only depends on $M_{\rm h}$.
The derivative of \eqref{new_areaderivative} with respect to $M_{\rm h}$ equals zero when $M_{\rm h} = \sqrt{M}$, which corresponds to a square-shaped array. However, this is the maximum because the derivative is positive for $1 \leq M_{\rm h} < \sqrt{M}$ and negative for $\sqrt{M} < M_{\rm h} \leq M$. Hence, the minimum total area is achieved at the boundary of the feasible region: $M_{\rm h} = M$ and $M_{\rm v}=1$ or vice versa. 

We showcase the minimization of the total aperture area with respect to the number of horizontal/vertical antennas in Fig. \ref{aperturearea}. We can see that the total area is maximized when we have a square-like array, which becomes exactly a square when $\alpha=\gamma=0.5$. Conversely, the total area is minimized when the array configuration is a horizontal or vertical ULA. This result holds regardless of the values of $\alpha$ and $\gamma$.

In summary, we have proved the following main result.

\begin{Theorem}
Among all transmitter and receiver array configurations that maximizes the capacity by satisfying the condition \eqref{eq:8}, it holds that:
\begin{enumerate}
    \item The total aperture length is minimized by having USAs with $M_{\rm h} = M_{\rm v} = \sqrt{M}$ and equal antenna spacing:
\begin{equation}\label{eq:9b}
    {\rm  h}_{\rm t}= {\rm  h}_{\rm r} = {\rm  v}_{\rm t}= {\rm  v}_{\rm r}\sqrt{\frac{\lambda d}{\sqrt{M}}}.
\end{equation}
    \item The total aperture area is minimized by having horizontal ULAs with $M_{\rm h} = M$ and $M_{\rm v} = 1$ and equal antenna spacing
\begin{equation}\label{eq:9b_ULA}
    {\rm  h}_{\rm t}= {\rm  h}_{\rm r} = {\rm  v}_{\rm r}\sqrt{\frac{\lambda d}{M}}.
\end{equation}
Alternatively, vertical ULAs with equal spacing can be used.
\end{enumerate}
\end{Theorem}

Depending on which design metric is considered, we obtain two very different optimal configurations. Looking at contemporary systems, we notice that ULAs are mainly used when the antenna number is small, while URAs/USAs dominate when deploying MIMO systems with many antennas.
Hence, we conclude that having the smallest possible aperture length is of the utmost importance in future systems, and we will consider the USA configuration in the remainder of the paper.

\section{Capacity versus Carrier Frequency}\label{Sect_Capacity_Freq}

There is a progressive trend toward embracing higher carrier frequencies (i.e., smaller wavelengths) in future wireless communication systems. This development is usually motivated by uncovering larger available bandwidths that can enhance the system's capacity. Since the physical size of an antenna shrinks when the frequency increases (if the gain is maintained), we can also incorporate a greater number of antennas within the same physical array area. This is also necessary to compensate for the reduced area per antenna through increased beamforming gains. 
In this section, we will investigate how an increase in carrier frequency affects the optimal MIMO channel capacity if the total area of the transmitter and receiver is fixed. The goal is to explore if we can make efficient use of a higher multiplexing gain when the carrier frequency increases.

\subsection{Theoretical Capacity Analysis}

We consider identical USAs at the transmitter and receiver where the number of horizontal and vertical antennas are equal: $M_{\rm h} = M_{\rm v}=\sqrt{M}$. The optimal horizontal/vertical antenna spacing is $\sqrt{\frac{\lambda d}{\sqrt{M}}}$ as given in \eqref{eq:9b}. 
As discussed in Section~\ref{subsec:optimal_capacity}, the capacity expression with the ideal XPD ($\kappa=0$) can be written as 
\begin{equation}\label{capacityusa}
    C= 2B M \log_2 \bigg(1+\underbrace{\frac{P  \beta  }{2 B N_0}}_{\rm SNR} \bigg) \quad \text{bit/s},
\end{equation}
where $B$ is the bandwidth of the system and $\sigma^2 = B N_0$ exposes how the noise variance depends on the bandwidth. The channel gain $\beta = G^{\rm t} G^{\rm r} \left(\frac{\lambda}{4 \pi d} \right)^2 $ depends on the wavelength $\lambda$.
However, it is worth noting that the SNR term in \eqref{capacityusa} is independent of the number of antennas. On the one hand, the transmit power is divided over $2M$ antennas. On the other hand, there is a receive beamforming gain of $M$, so these effects cancel out apart from the $1/2$ term in \eqref{capacityusa}.

If the area of the array is fixed, the number of antennas ($M$) that can be deployed with the optimal spacing is determined by various factors, including the wavelength and the width of the antenna elements. In the following lemma, we provide an analytical expression for the number of antennas in this case.

\begin{Lemma}\label{Lem_num_of_antennas}
    Suppose the array areas $A=A_{\rm t}=A_{\rm r}$ are fixed and each antenna has the width/height $W=w \lambda$ for some constant $w \geq 0$. The maximum number of antennas that can be deployed in a USA while following the optimal antenna spacing in \eqref{eq:symmetric-separation} is
\begin{equation}\label{eq_M_wrt_A}
    M= \left(\frac{k_0 + \sqrt{k_0^2-4}}{2} \right)^2,
\end{equation}
where $k_0 = 2 + \frac{1}{\lambda d} (w\lambda - \sqrt{A})^2  $. 
\end{Lemma}
\begin{IEEEproof}
    The proof is given in Appendix~\ref{proof_eq_M_wrt_A}.
\end{IEEEproof}

The expression in Lemma~\ref{Lem_num_of_antennas} depends on the size of the outermost antennas, but when the carrier frequency is large (i.e., the wavelength is small), we obtain the following approximation.

\begin{Lemma}\label{Approximation_of_M}
When $\lambda$ is small, we can approximate $M$ in \eqref{eq_M_wrt_A} by $(\frac{A}{\lambda d})^2$ since
\begin{equation}\label{eq:lim_M}
 {\rm lim}_{\lambda \rightarrow 0} \frac{M}{(A/(\lambda d))^2} = {\rm lim}_{\lambda \rightarrow 0}\left( \frac{k_0 + \sqrt{k_0^2-4} }{2 A / (\lambda d)}\right)^2 =1,
\end{equation}
where $k_0$ was defined in Lemma \ref{Lem_num_of_antennas}.
\end{Lemma}

\begin{IEEEproof}
     We expand $k_0 = 2+ \frac{w^2 \lambda}{ d} - \frac{2w\sqrt{A}}{d} + \frac{A}{\lambda d} $ and substitute it into \eqref{eq:lim_M}. It then follows that the limit is $1$.
\end{IEEEproof}

\begin{figure}
    \centering
    \includegraphics[width=1\textwidth]{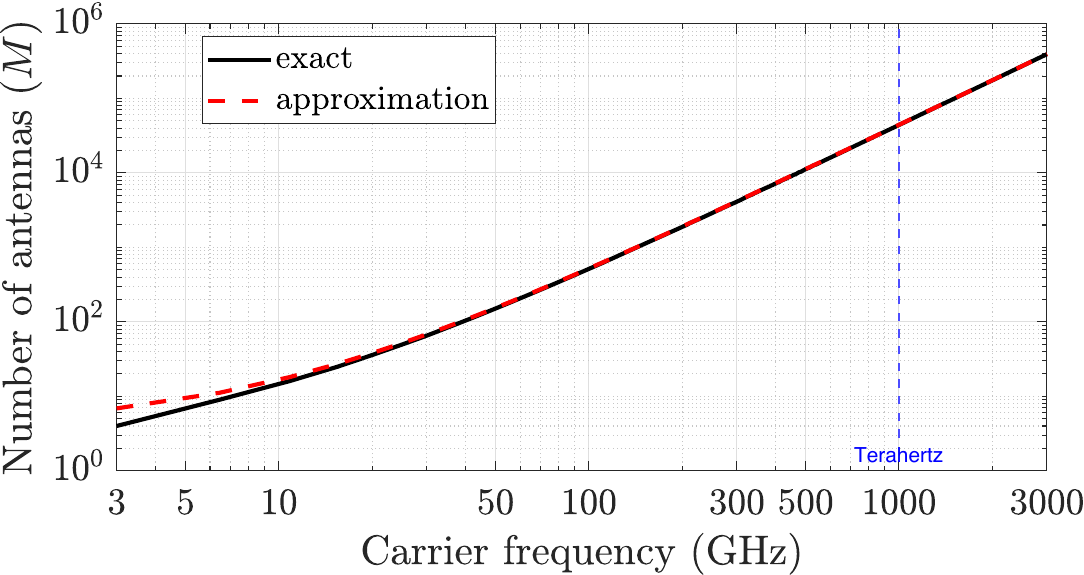}
    \caption{The number of antennas with respect to the carrier frequency. We consider $A=5 \text{ m}^2$, $W=\lambda/2$, and $d=80$ m.}
    \label{fig_M}
\end{figure}

We investigate the exact expression in \eqref{eq_M_wrt_A} by plotting the number of antennas with respect to the wavelength in Fig.~\ref{fig_M} for $A=5 \text{ m}^2$, $W=\lambda/2$, and $d=80$ m. We can see that as the carrier frequency increases (i.e., the wavelength shrinks),  the number of antennas that fits into the fixed area increases. We also show the approximation from Lemma~\ref{Approximation_of_M} in the figure. There is a slight approximation error when $f_c < 10$ GHz, but for a higher carrier frequency corresponding to $\lambda \leq 0.01$, the approximation is highly accurate, as expected since it is exact in the asymptotic limit.
Hence, at mmWave frequencies and beyond, the number of antennas is approximately proportional to $1/\lambda^2$. Hence, by increasing the carrier frequency by a factor of $10$, we can not only deploy $100$ more antennas in the same area (which is obvious) but achieve a channel with $100$ times more equal-sized singular values in the channel matrix.

We will further analyze how the capacity expression in \eqref{capacityusa} depends on the carrier frequency $f_c$. We consider arrays with the area $A = 5 \text{ m}^2$ that communicate over the distance $d=80$ m. This could be an outdoor LOS link for backhaul or fixed wireless access.
The transmit SNR is fixed at $P/N_0=204$ dB and $W = \lambda/2$.
When varying the carrier frequency, we consider two different cases. In Fig.~\ref{fig_C_vs_lambda}(a), the bandwidth is proportional to the carrier frequency such that $B = 0.03 f_c$. This is roughly the case in current systems (e.g., $90$ MHz at $3$\,GHz or $900$\,MHz at $30$ GHz). In Fig.~\ref{fig_C_vs_lambda}(b), we instead keep the bandwidth constant at $B=90$ MHz, so the two subfigures begin at the same capacities at $f_c=3$\,GHz.
Three different antenna designs are compared in Fig.~\ref{fig_C_vs_lambda}:

\begin{figure*}
\centering
\subfloat[Capacity vs. carrier frequency with varying bandwidth ($B= 0.03 f_c$ Hz)]
{\includegraphics[width=0.49\textwidth]{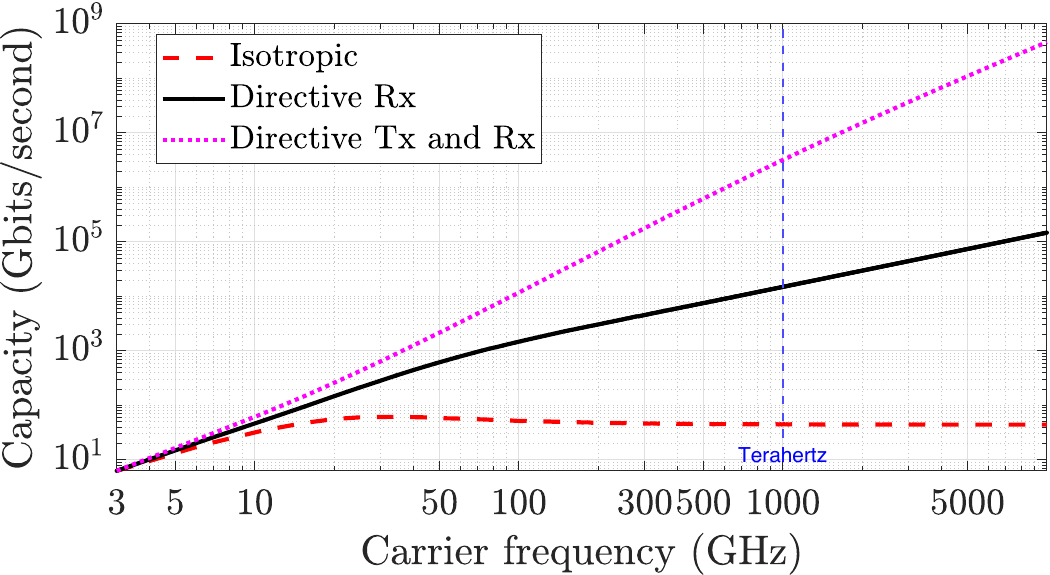}}\hfill
\centering
\subfloat[Capacity vs. carrier frequency with constant bandwidth ($B=90$ MHz)]
{\includegraphics[width=0.49\textwidth]{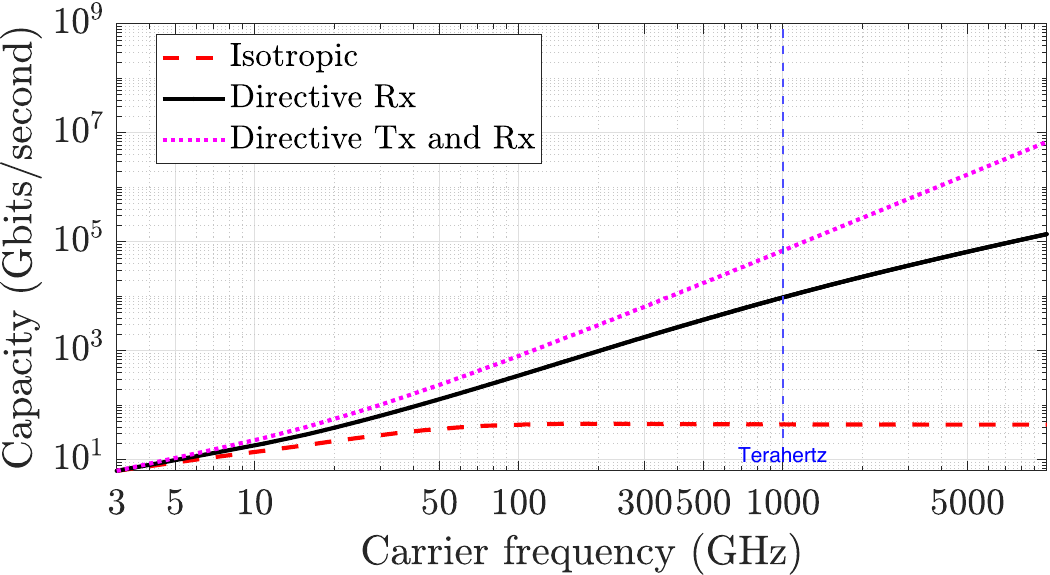}}\hfill
\centering
\caption{The graph between capacity and wavelength. In the case of isotropic antennas, $G^{\rm t} = G^{\rm r} = 1$; In the case of directive Rx, $G^{\rm t} = 1$ and $G^{\rm r} = 1/\lambda$; In the case of directive Tx and Rx, $G^{\rm t} = G^{\rm r} = 1/\lambda$. We set $A = 5\text{ m}^2$, $P/N_0=204$ dB, $d=80$ m, and $c = 1/2$.}
\label{fig_C_vs_lambda}
\end{figure*}

\begin{itemize}
    \item \textit{Isotropic transmit antenna (Tx) and receive antenna (Rx).} In this case, we keep the antenna gains fixed at $G^{\rm t}=G^{\rm r}=1$ when varying the carrier frequency. Therefore, $\beta$ is proportional to $\lambda^2$. The capacity with respect to the carrier frequency
    is illustrated by the red dashed curve in Fig.~\ref{fig_C_vs_lambda}. The capacity grows with the frequency since both the bandwidth $B$ and multiplexing gain $M$ increase in Fig.~\ref{fig_C_vs_lambda}(a), while only the latter increases in Fig.~\ref{fig_C_vs_lambda}(b). However, the SNR gradually reduces because $\beta$ shrinks as $f_c$ increases. This SNR reduction eventually eliminates the benefits of performing spatial multiplexing and, therefore, the capacity converges to an upper limit. Interestingly, the same capacity limit is attained in (a) and (b), but we reach it more rapidly in the case where the bandwidth increases since the SNR decays faster when the power is divided over more spectrum.

    \item \textit{Isotropic Tx  and directive Rx.} 
    It is not the propagation loss that leads to a reduced SNR when $f_c$ increases, but the assumption of isotropic receive antennas whose effective area is proportional to $\lambda^2$ and captures less power at higher frequencies.
    To avoid this effect, we can utilize directive receive antennas to compensate for the SNR reduction caused by the shrinking wavelength. The green curves in Fig.~\ref{fig_C_vs_lambda} depict the capacity when we set $G^{\rm r} = 1/\lambda$ while keeping $G^{\rm t}=1$.\footnote{This can be achieved by having aperture antennas at the receiver whose physical dimensions $a$ and $b$ are selected as $a=\gamma\lambda$ and $b=\gamma\lambda$, where $\gamma$ is proportional to $1/\sqrt{\lambda}$ so that the area $ab$ becomes proportional to $\lambda$. Alternatively, each antenna can consist of $N_e$ isotropic elements, where $N_e$ is proportional to $\sqrt{\lambda}$ so that the total area becomes proportional to $1/\sqrt{\lambda}$.} In this case, the SNR reduction is partially compensated, since $\beta$ is proportional to $\lambda$ instead of $\lambda^2$. Hence, when $f_c$ grows large, the capacity improves significantly compared to the case of having isotropic transmit and receive antennas, because the multiplexing gain now increases more rapidly than the SNR decays.
    For instance, for a carrier frequency of $f_c = 30$ GHz in Fig.~\ref{fig_C_vs_lambda}(a), the capacity improves by approximately an order of magnitude compared to the isotropic case. Moreover, there is no asymptotic limit.
    
    \item \textit{Directive Tx and Rx.} To further improve the capacity, we can use directive antennas both in the transmitter and receiver, which can be configured with $G^{\rm t}=G^{\rm r}=1/\lambda$ (the same antenna design can be used as in the previous case). The black curves in Fig.~\ref{fig_C_vs_lambda} depict the corresponding capacity. 
    Compared to the case of isotropic Tx and directive Rx, we now see an even faster capacity growth because the SNR is independent of the wavelength due to selected antenna gains. We can achieve enormous capacities thanks to the vast number of spatial and spectral DOF obtained as $f_c$ increases. As seen from Fig.~\ref{fig_C_vs_lambda}(b), we don't even need more spectrum to reach a massive capacity, but the multiplexing gain is sufficient.
\end{itemize}

We observed that with isotropic antennas, there is a finite asymptotic limit in Fig.~\ref{fig_C_vs_lambda}(a) and (b). It is the same regardless of whether the bandwidth increases with $f_c$ or is fixed, and can be characterized as follows.

\begin{Theorem}\label{Approximation_of_C}
The MIMO capacity with isotropic transmit and receive antennas has the finite limit
\begin{equation} \label{eq:ClimitIso}
C \to \left(\frac{A}{ 4 \pi d^2} \right)^2 \frac{P  }{N_0} \log_2(e) \quad \textrm{as } \lambda \to 0,
\end{equation}
regardless of whether the bandwidth is constant or inversely proportional to the wavelength $\lambda$.
\end{Theorem}
\begin{IEEEproof}
    Considering the case with constant bandwidth $B$, we have $\beta$ as given in \eqref{eq:first-approx}. Substituting it into   
 \eqref{capacityusa} gives us $C =  2 B M \log_2 (1+ \frac{P  \lambda^2  }{ 2 B N_0\left(4 \pi d\right)^2 }) $,
 where the SNR term goes to zero as $\lambda \to 0$.
 In the same asymptotic regime, $M $ can be approximated by $\left(\frac{A}{\lambda d}\right)^2$, according to Lemma~\ref{Approximation_of_M}. Hence, when $\lambda$ approaches zero, we can approximate the capacity as
 \begin{align}
 \notag
   C &= 2 B M \log_2 \left( 1+ \frac{P  \lambda^2  }{ 2 B N_0\left(4 \pi d\right)^2 }\right), \\\nonumber
   &\approx 2B\left(\frac{A}{\lambda d}\right)^2 \log_2(e) \left(\frac{P  \lambda^2  }{ 2 B N_0 \left(4 \pi d\right)^2 }\right)\\ 
   &= \left(\frac{A}{ 4 \pi d^2} \right)^2 \frac{P  }{N_0} \log_2(e)\label{Climit},
 \end{align}
 where we utilize the first-order Taylor expansion to approximate $\log_2(1+x) \approx \log_2(e) x$ that is tight as $t \to 0$. The approximation error vanishes as $\lambda \to 0$, which establishes the result in \eqref{eq:ClimitIso}.

Suppose the bandwidth is instead wavelength-dependent as $B = B_0 / \lambda$ for some constant $B_0$. The SNR will still approach zero as $\lambda \to 0$, which implies that the $B$-term in the pre-log factor and SNR will cancel asymptotically and we still achieve the limit in \eqref{Climit}.
This completes the proof.
\end{IEEEproof}

Although this theorem proves that the same asymptotic limit is approached regardless of the amount of spectrum, the convergence to the limit in \eqref{eq:ClimitIso} occurs at more modest carrier frequencies when the bandwidth grows when the frequency increases (i.e., the wavelength shrinks). We observed this in Fig.~\ref{fig_C_vs_lambda}.
The situation is much different when directive antennas are utilized at either side of the communication link, which can be formalized as follows.

\begin{corollary}
    If $G^{\rm t}G^{\rm r}=G_0/\lambda^{\rho}$ for some $\rho \in (0,2]$, then the capacity grows to infinity proportionally to $1/\lambda^{\rho}$ as $\lambda \to 0$.
\end{corollary}
\begin{IEEEproof}
When $\lambda \to 0$ and $\rho \in (0,2)$, it follows similar to \eqref{Climit} that $C = 2B\left(\frac{A}{\lambda d}\right)^2 \log_2(e) (\frac{P  \lambda^2 G_0 }{\lambda^{\rho} 2 B N_0 \left(4 \pi d\right)^2 })$, which is proportional to $1/\lambda^{\rho}$. When $\rho=2$, the SNR is wavelength-independent, but the pre-log factor is proportional to $1/\lambda^2$.
\end{IEEEproof}

This result implies that we can achieve any desired capacity value by increasing the carrier frequency, without requiring more bandwidth or power. We observed this already in Fig.~\ref{fig_C_vs_lambda}, where the magenta curves grow without bound as $O(f_c^2)$ because $\rho=2$, while the black curves grow as $O(f_c)$ because $\rho=1$.
Note that this is achieved while shrinking the antenna areas, which are proportional to $\lambda$ instead of $\lambda^2$ (as with isotropic antennas).
The convergence to the asymptotic behavior is a lot quicker when the bandwidth also grows with $f_c$.

\subsection{Realistic Antenna Modeling}

\begin{figure}
    \centering
    \includegraphics[width=1\textwidth]{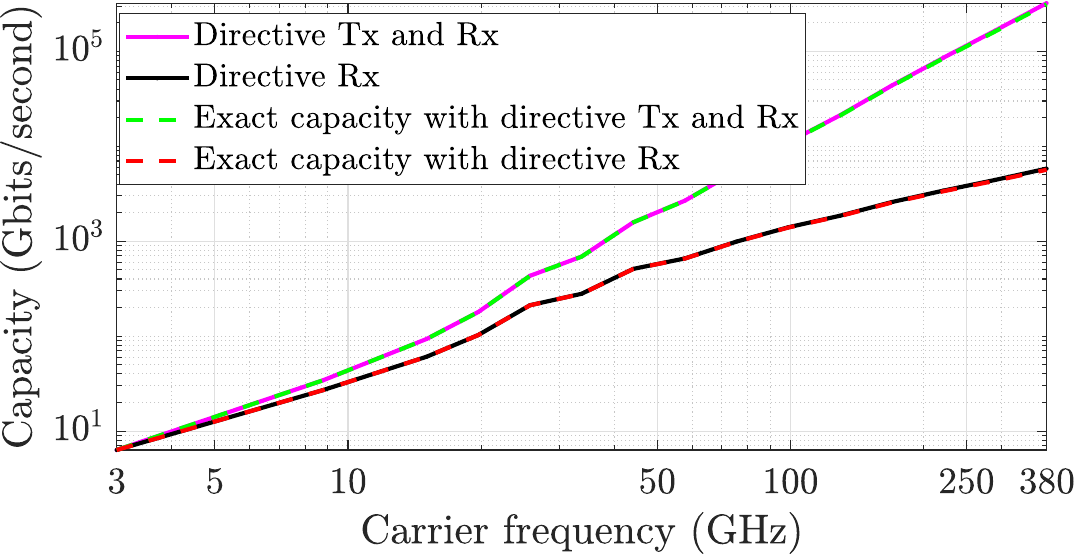}
    \caption{Capacity comparison with ideal and realistic antenna gain models  with $A=5 \text{ m}^2$, $W=\lambda/2$, $d=80$ m, and $B= 0.03 f_c$ Hz.}
    \label{fig_Capa_comparison}
\end{figure}

The directive antennas considered in the previous simulations are idealized in the sense of providing maximum gain to all antennas in the other array. 
In this section, we validate those results by considering a more realistic antenna gain model, which includes directivity and polarization losses \cite{2020_Björnson_JCommSoc}. The general gain of an aperture antenna located in the $xy$-plane can be calculated by as \cite{Kay1960a}
\begin{equation}\label{eq_II_NormalizedAntennaGain}
   G = \frac{\left| \int_{\mathcal{A}} E(x,y) dx dy \right|^2}{A_{\rm phy} \int_{\mathcal{A}} \left|E(x,y)\right|^2 dx dy },
\end{equation}
where $A_{\rm phy}$ is the antenna's physical area, $\mathcal{A} \subset \mathbb{R}^2$ is the range of $xy$-coordinates spanned by the antenna, where $\mathcal{A}=\left\{(x, y):|x|\leq \frac{A_{\rm phy}}{2},|y|\leq\frac{A_{\rm phy}}{2}\right\}$ and $ E (x,y)$ is the incident electrical field.
If the transmitter is located at the coordinate $(\bar{x},\bar{y},\bar{z})$, the electric field at point $(x,y)$ can be expressed as \cite{2020_Björnson_JCommSoc,Decarli2021a}
\begin{equation}\label{eq_II_ElectField}
 E (x,y) =  \frac{E_0}{\sqrt{4 \pi}} \frac{\sqrt{z (\left( x-\bar{x} \right)^2+\bar{z}^2)}} 
 { r ^{5/4}}   e^{-j\frac{2\pi}{\lambda} \sqrt{r}},
\end{equation}
where $E_0$ is the electric intensity of the transmitted signal (in Volt) and $r =  ( ( x-\bar{x})^2 + (y-\bar{y})^2 + \bar{z}^2 )$ denotes the squared Euclidean distance between the transmit and receive antennas. By substituting \eqref{eq_II_ElectField} into \eqref{eq_II_NormalizedAntennaGain} and considering square-shaped antennas, we can compute the transmitter gain $G_{m,k}^t$ and receiver gain $G_{m,k}^r$ between any pair of antennas.
By substituting these values into \eqref{beta_m_k}, we obtain the corresponding $\beta_{m,k}$ and can calculate the exact MIMO channel in \eqref{eq:Hu}, compute the optimal power allocation, and finally the capacity. When using a directive antenna, we reduce the physical area of the transmitter/receiver antenna by a factor of $\lambda$ when solving the integrals. 

In Fig.~\ref{fig_Capa_comparison}, we show the MIMO capacity for carrier frequencies up to 380 GHz. The frequency range is smaller than in previous figures to limit the computational complexity but covers all practically interesting cases. An area of $5$ m$^2$ is considered for the transmitter and receiver arrays, the arrays are $d=80$\,m apart, and the bandwidth grows with the frequency as in previous figures.
The figure shows that the capacity achieved with the realistic antenna model is very close to that reported in previous figures.
This holds both when only the receiver array has directive antennas, and when there are directive antennas at both sides of the link.

\subsection{Implementation with Practical User Devices}

The previous simulation examples have considered point-to-point MIMO scenarios with equal-sized arrays at the transmitter and receiver; that is, $\alpha=\gamma=0.5$ using the antenna spacing definitions in \eqref{eq_decompose1} and \eqref{eq_decompose2}. This approach minimizes the total aperture area and is attractive for fixed wireless links.
However, when it comes to serving user devices, it is common that the base station array is significantly larger in size than the device.
By utilizing the parameters $\alpha$ and $\gamma$, we have the flexibility to tailor the antenna spacing in each array, while maintaining the maximum MIMO capacity. This flexibility enables us to configure the antenna elements in the transmitter array (base station) to be sparse and concurrently achieve a receiver array (device) with relatively small antenna spacing. 

Let us now consider a concrete numerical example where the base station and device are $d=70$ meters apart. Each USA consists of 128 antenna elements (i.e., $M_{\rm h}=M_{\rm v}=8$) with equal vertical and horizontal spacings ($\alpha=\gamma$) and $W=\lambda/2$.
When we set $\alpha=0.5$, we can attain a capacity of $3.2$ Tbps at the carrier frequency $100$ GHz with $B=3$ GHz and a transmit power of 1 W per antenna element. However, to give the device a more attractive size, we can select $\alpha = 0.01$, which makes the area of the receiver array equal to $0.0369$ m$^2$. This is feasible to integrate into a smartphone. Simultaneously, the area of the transmitter array becomes $45.57$ m$^2$, which can be deployed on a building or tower, given its sparse nature with $0.9642$\,m in both the horizontal and vertical directions. We achieve a capacity of 3.1 Tbps in this case, which is close to the theoretical upper bound. While some of this extraordinary number comes from the wide bandwidth, it is the spatial multiplexing of $128$ streams that makes the larger difference compared to contemporary technology.
\begin{figure}
    \centering
    \includegraphics[width=1\textwidth]{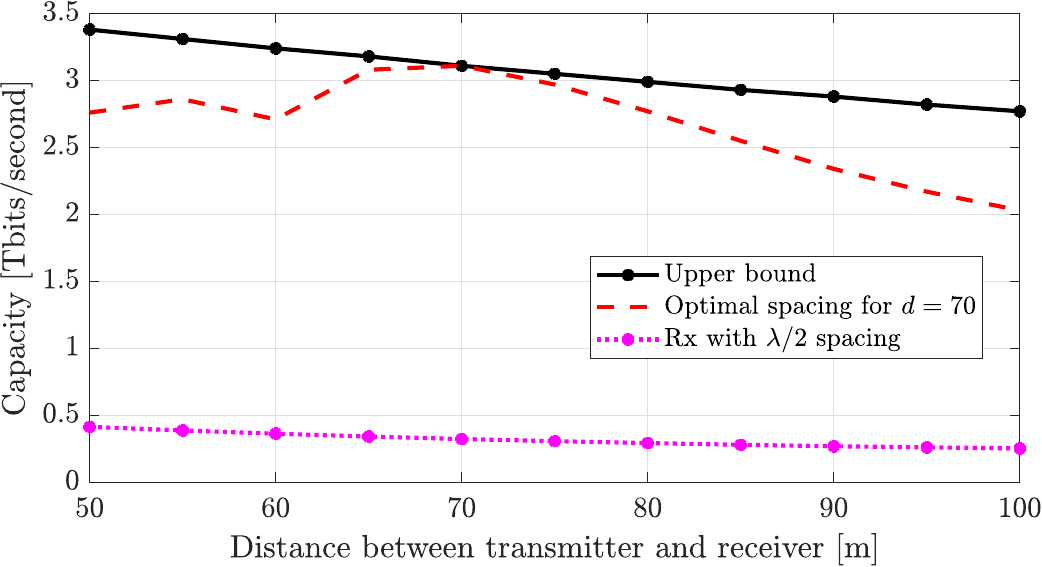}
    \caption{Capacity vs. distance between the transmitter and receiver with $\alpha=0.01$ and $M_{\rm h}=M_{\rm v}=8$ antennas at the transmitter and receiver. The upper bound has an optimal antenna spacing at every point, while the dashed curve considers a fixed antenna spacing designed for $70$\,m.}
    \label{fig_Capa_skyscrapperdistance}
\end{figure}

The optimal spacing was derived for a specific propagation distance $d$ between the transmitter and receiver. However, if the user is mobile, there could be a mismatch between $d$ and the actual propagation distance in practice, which leads to sub-optimal capacity. To investigate the implications of this, we plot the capacity with respect to the propagation distance in Fig.~\ref{fig_Capa_skyscrapperdistance}. The solid black curve shows the optimal capacity if the antenna spacing is adjusted for each distance.
 The dashed red curve shows the actual capacity when the antenna spacing is fixed and optimized for $d=70$ m. These curves only intersect at $70$\,m while there is otherwise a performance loss from having a suboptimal spacing. The capacity loss is 4.5$\%$ at $75$\,m compared to $70$\,m, but it is mostly due to the increased pathloss because the difference to the black curve is only 1.9$\%$ at $75\,$m. However, the capacity losses can be substantially larger when the propagation distance is much different from the one considered when selecting the antenna spacing.

\begin{figure}
    \centering
    \includegraphics[width=1\textwidth]{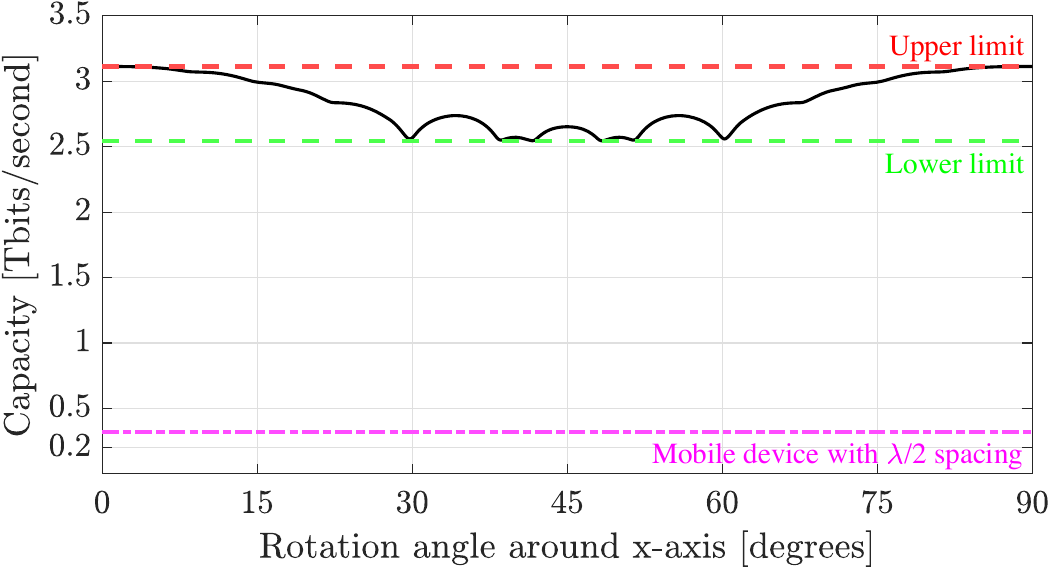}
    \caption{Capacity vs. rotation angle in the $xy$-plane with $\alpha=0.01$ and $M_{\rm h}=M_{\rm v}=8$ antennas at the transmitter and receiver.}
    \label{fig_Capa_skyscrapperrotation}
\end{figure}

In practice, the mobile device might not be aligned perfectly in the broadside direction of the array. In Fig.~\ref{fig_Capa_skyscrapperrotation}, we plot the capacity with respect to the rotation angle in the $xy$-plane where the array is deployed. As expected, the capacity fluctuates and generally reduces when the rotation angle is different. This is because the rotation makes the antenna spacing look smaller from the base station's perspective. We observe that the lowest capacity is achieved when the device is rotated around $60$ degrees. This lower limit capacity is $17\%$ below the optimal capacity achieved with broadside arrays. Nevertheless, the capacity is still 7.87 times larger than with a conventional $\lambda/2$-spaced array, which is shown as a reference curve. Hence, it is worth using widely spaced antenna arrays even if we cannot always reach the maximum capacity for mobile users.

\section{Conclusion}\label{Sect_conclusion}

In this paper, a LOS MIMO channel matrix with equal singular values is obtained by identifying the optimal horizontal and vertical antenna spacings for a rectangular array.
We considered dual-polarized antennas with imperfect XPD and noticed that the XPD affects the capacity value and precoding, but not optimal antenna spacing. The capacity with dual-polarized antennas is always higher than that with the same number of single-polarized antennas. The optimal configuration makes use of radiative near-field properties and focuses signals into tiny beams with adjacent antennas at the null angles.
There are many rectangular array arrangements that achieve the maximum capacity; thus, we considered minimizing the physical dimensions of the array in terms of either aperture length or aperture area. We proved analytically that square arrays exhibit the minimum aperture length and are preferred at deployment locations with a specific maximum dimension. By contrast, linear arrays have the minimum aperture area and are preferred when withstanding the wind load is the main limiting factor.

The capacity of a communication link with fixed aperture areas at the transmitter and receiver can be increased by using more bandwidth or antennas. Both approaches are tightly connected with moving towards higher carrier frequencies $f_c$.
We proved analytically that the MIMO rank is proportional to $f_c^2$. Hence, spatial multiplexing provides a more rapid capacity growth than the available bandwidth, which is roughly proportional to $f_c$.
Hence, it is more important to move to higher frequencies to get ``more MIMO'' than more bandwidth.
However, the spatial multiplexing gain is only helpful if we can maintain a high SNR when increasing the frequency because the power must be spread over more signal dimensions.
This can be achieved using weakly directive antennas with an effective area that shrinks proportional to $\lambda$ instead of $\lambda^2$ (as with fixed-gain antennas).

The numerical results demonstrate that massive capacities beyond 1 Tbps can be reached at conventional mmWave frequencies (30-100 GHz). These results are directly useful for wireless backhaul links and fixed wireless access with equal-sized arrays at the transmitter and receiver. The analytical results also give the flexibility to tailor the antenna spacing in each array, thus, Tbps rates can be maintained in mobile access scenarios with a sparse transmitter array (base station) and a closely spaced receiver array (mobile device). It appears that spatial multiplexing through well-designed MIMO arrays can enable massive capacity links without requiring sub-THz spectrum, even if the arrays are not perfectly aligned in angle and at the optimal distance.

High-rank MIMO channels can also be achieved over long distances. If the carrier frequency is 75 GHz (E-band) and propagation distance is 1 km, then we can deploy $4\times4$ dual polarized square arrays with an optimal antenna spacing of around $1$ m in a fixed communication link.

The results in this paper can be directly applied to multi-user scenarios with orthogonal scheduling. The extension to multi-user MIMO communications is interesting. We believe that large antenna spacing will remain important to achieve high-rank channels for each user; however, other optimization techniques are required to attain optimal antenna spacing. The methods recently developed for moveable antennas in  \cite{zhu2024}, \cite{moveable23} and for tri-polarized antennas in \cite{multiusertri} might be useful when tackling this problem.

\appendices

\section{Derivation of \eqref{eq_M_wrt_A}}
\label{proof_eq_M_wrt_A}

We consider USAs with equal spacing in the horizontal and vertical dimensions ${\rm  h}_{\rm  t} = {\rm  v}_{\rm t} = \sqrt{\frac{\lambda d}{\sqrt{M}}}$, where $\sqrt{M} = M_{\rm h} = M_{\rm v}$. We can calculate the area of the array as
 \begin{align}\nonumber
     A &= \left( \sqrt{\frac{\lambda d}{\sqrt{M}}}  (\sqrt{M}-1)+W \right) \left( \sqrt{\frac{\lambda d}{\sqrt{M}}}  (\sqrt{M}-1)+W  \right),\\ 
       &=(\lambda d) x^2+2w\sqrt{\lambda^3 d}x+(w\lambda)^2,
       \label{fixedarea}
 \end{align}
where $x= \sqrt[4]{M}-\frac{1}{\sqrt[4]{M}}$ and $W=w\lambda$ with $w \geq 0$ being a constant. By utilizing the classical quadratic root formula for quadratic expression in \eqref{fixedarea}, we obtain $x=-w\sqrt{\frac{\lambda}{d}}+\sqrt{\frac{A}{\lambda d}} $.
We can further express the relation between $M$ and $\lambda$ as
\begin{equation}\label{eq_B}
    \sqrt[4]{M}-\frac{1}{\sqrt[4]{M}}  +w\sqrt{\frac{\lambda}{d}}-\sqrt{\frac{A}{\lambda d}} =0
\end{equation}
By solving \eqref{eq_B} for $M$, we obtain the final expression in \eqref{eq_M_wrt_A}.

\bibliographystyle{IEEEtran}
\bibliography{IEEEabrv,mybib}

\end{document}